\documentclass{IEEEtran}
\usepackage{cite}
\usepackage{amsmath,amssymb,amsfonts}
\usepackage{algorithmic}
\usepackage{graphicx}
\usepackage{textcomp}
 \usepackage[caption=false,font=footnotesize]{subfig}
 \usepackage{placeins}
 \usepackage{color}
\def\BibTeX{{\rm B\kern-.05em{\sc i\kern-.025em b}\kern-.08em
    T\kern-.1667em\lower.7ex\hbox{E}\kern-.125emX}}

\newcommand{\Xee}{\chi_\text{ee}}
\newcommand{\Xmm}{\chi_\text{mm}}
\newcommand{\Xem}{\chi_\text{em}}
\newcommand{\Xme}{\chi_\text{me}}

\newcommand{\A}{\mathbf{A}}

\newcommand{\xbold}{\mathbf{x}}

\newcommand{\fbold}{\mathbf{f}}

\newcommand{\CTwo}{\mathcal{C}_2}

\newcommand{\norm}[1]{\left\lVert#1\right\rVert}

\newcommand{\Avone}{\mathbf{A}_{\xi_1}}
\newcommand{\Avtwo}{\mathbf{A}_{\xi_2}}
\newcommand{\nvec}{\mathbf{f}}

\newcommand{\Wmat}{\mathbf{W}}

\newcommand{\Axpol}{\mathbf{A}_{\text{X-pol}}}

\newcommand{\designfxnaloneRHS}{
	\left\|  {\Wmat\left( \left|\Avone \xbold\right|^2 + \left|\Avtwo\xbold\right|^2 \right)} - \Wmat\nvec \right\|^{2}_{2}
}
\newcommand{\designfxnaltwo}{
	\CTwo\left(\xbold \right) =  \left\| \Axpol \xbold \right\|^{2}_{2}
}

\makeatletter
\def\ps@IEEEtitlepagestyle{%
  \def\@oddfoot{\mycopyrightnotice}%
  \def\@evenfoot{}%
}
\def\mycopyrightnotice{%
  {\begin{minipage}{\textwidth}\ \\[10pt]
  \footnotesize \copyright2019 IEEE. Personal use of this material is permitted. Permission from IEEE must be obtained for all other uses, in any current or future media, including reprinting/republishing this material for advertising or promotional purposes, creating new collective works, for resale or redistribution to servers or lists, or reuse of any copyrighted component of this work in other works.\hfill\end{minipage}
}
  \gdef\mycopyrightnotice{}
}

\begin{document}
\title{On the Use of Electromagnetic Inversion\\for Metasurface Design}
\author{Trevor Brown, \IEEEmembership{Student Member, IEEE}, Chaitanya Narendra, \IEEEmembership{Student Member, IEEE}, Yousef Vahabzadeh, \IEEEmembership{Student Member, IEEE}, Christophe Caloz, \IEEEmembership{Fellow, IEEE} and Puyan Mojabi, \IEEEmembership{Member, IEEE}
\thanks{}
\thanks{T. Brown, C. Narendra, and P. Mojabi are with the Department of Electrical and Computer Engineering, University of Manitoba, Winnipeg, MB, Canada (e-mail: umbrow47@myumanitoba.ca). }
\thanks{Y. Vahabzadeh and C. Caloz are with the Department of Electrical Engineering, \'Ecole Polytechnique de Montr\'eal, Montr\'eal, QC, Canada (e-mail: christophe.caloz@polymtl.ca).}
}

\maketitle

\begin{abstract}
We show that the use of the electromagnetic inverse source framework offers great flexibility in the design of metasurfaces. In particular, this approach is advantageous for antenna design applications where the goal is often to satisfy a set of performance criteria such as half power beamwidths and null directions, rather than satisfying a fully-known complex field. In addition, the inverse source formulation allows the metasurface and the region over which the desired field specifications are provided to be of arbitrary shape. Some of the main challenges in solving this inverse source problem, such as formulating and optimizing a nonlinear cost functional, are addressed. Lastly, some two-dimensional (2D) and three-dimensional (3D) simulated examples are presented to demonstrate the method, followed by a discussion of the method's current limitations.
\end{abstract}


\section{Introduction}
\label{sec:introduction}
\IEEEPARstart{E}{lectromagnetic} metasurfaces are structures of subwavelength thickness that are composed of a subwavelength lattice of scattering elements~\cite{holloway2012overview, PfeifferGrbicMetasurface,SelvanayagamOpticsExpress,tretyakov2015metasurfaces,MaciSpace2015,CalozReview2018}. In particular, transmitting metasurfaces, which are the focus of this paper, offer a systematic transformation of an incident field (input field) to a desired transmitted field (output field) by imposing appropriate surface boundary conditions. From a circuit view point, these surfaces may be thought of as collections of several two-port networks transforming the input field to the output field~\cite{EleftheriadesCircuitModelMetasurface}. In other words, a metasurface may be considered as an electromagnetic wave transformer \cite{CalozWaveTransformer}:~similar to electric transformers that enable the \emph{systematic} change of a given voltage level to a desired one, electromagnetic metasurfaces have the potential to enable the \emph{systematic} transformation of a given electromagnetic field to a desired one. Thus, they can play an important role in antenna engineering applications.

Recently, a few metasurface design methods have been presented, e.g., see \cite{epstein2016huygens, achouri2015general}. In particular, two main design steps are outlined in \cite{epstein2016huygens}: macroscopic and microscopic. The macroscopic step finds the surface electric/magnetic impedance/admittance profiles, or equivalently, the electric and magnetic susceptibility profiles, required for the metasurface to transform the input field into the output field. In contrast, the microscopic step is concerned with the physical implementation of this surface impedance profile with the design of appropriate subwavelength elements such as a triple-dogbone topology~\cite{8741180}. The macroscopic aspect is the focus of this work, and therefore the physical implementation of subwavelength elements are not considered in this paper. 

To the best of our knowledge, all these synthesis methods rely on the knowledge of the tangential fields on the input and output surfaces of the metasurface. However, in some applications these tangential fields are not readily available. For example, consider the case where the designer would like to transform the radiation pattern of a horn antenna to a different radiation pattern with the following design specifications: main beam direction, half-power-beamwidth (HPBW), and side lobe levels (SLL). The knowledge of these design specifications does not directly translate into the knowledge of a tangential field on the output surface of the metasurface. This case clearly presents a limitation for the existing metasurface synthesis methods. This is important since many antenna design applications rely on these types of specifications, and not on an explicit expression of a desired complex field. 

\subsection{Electromagnetic Inversion}
As can be seen from the discussion above, there is a need for greater flexibility in the design of metasurfaces, and in particular the ability to design from a set of desired performance criteria (e.g., HPBW, null directions, etc.) is necessary. This paper proposes to cast the metasurface design problem as an electromagnetic inverse source problem. The main goal in electromagnetic inverse problems is to find an unknown \emph{cause} from its known \emph{effect}~\cite{NikolovaBook}. For example, the known effect could be a desired HPBW, and the cause to be found could be a set of electric and magnetic currents that generate a radiation pattern exhibiting such a HPBW. (Once these currents are found, the surface properties of the metasurface can be obtained.) The act of processing the effect to find its cause is often referred to as \emph{inversion}, and the obtained cause is often referred to as the \textit{reconstructed} result (e.g., the reconstructed electric current). Broadly speaking, electromagnetic inversion can be classified into two categories. The first category is inverse \textit{scattering}, in which the goal is to reconstruct complex permittivity and/or permeability profiles that produce a known scattered field when illuminated with a known incident field. The second category, which is the one used in this work, is inverse \textit{source}, in which the goal is to reconstruct a set of electric and/or magnetic currents that generate a known electromagnetic signature. 

\subsection{Electromagnetic Inversion for Chacterization}
There are various applications for electromagnetic inversion within a wide frequency spectrum (from a few hertz \cite{AriaMRGNI} to optical frequencies~\cite{AriaOptical}), most of which relate to some form of \textit{characterization}. In such applications, electromagnetic inverse scattering or inverse source algorithms are applied to \textit{measured} data. For example, in antenna diagnostics~\cite{lopez2009improved,brown2018CAMA}, the near-field data of an antenna under test are measured. From this measured data, the equivalent currents of the antenna can be reconstructed using an electromagnetic inverse source algorithm, and can then, for example, be used for finding faulty elements in an antenna array. As another example, in microwave breast imaging~\cite{MeaneyBreast2012}, the breast is illuminated from different directions, and the resulting scattered fields are measured. From these measured scattered data, the complex permittivity profile of the breast can be reconstructed using an electromagnetic inverse scattering algorithm, and can then be potentially used for detecting tumours. One of the main advantages of using the electromagnetic inversion framework is that it can work with various forms of data. For example, electromagnetic inversion algorithms can invert phaseless (amplitude-only) near-field measurements \cite{brown2017multiplicatively}. In addition, they are not merely limited to canonical measurement surfaces (e.g., planar, cylindrical, spherical), and work with arbitrarily shaped measurement domains~\cite{ChaiANTEM2018}. Moreover, electromagnetic inversion algorithms can systematically incorporate prior information into their formulation to enhance the resulting reconstruction accuracy~\cite{MeaneyMRI2019}.

\subsection{Electromagnetic Inversion for Design}
As noted above, the electromagnetic inversion framework has been mainly used for characterization applications. To adapt it to design applications, we mainly need to replace the \textit{measured} data with \textit{desired} data in the mathematical formulation, and then reconstruct a cause that can generate the desired effect. For example, in~\cite{bucci2005synthesis}, it was suggested that we can use electromagnetic inverse scattering algorithms for the design of dielectric profile lenses. Inspired by this work, inverse scattering algorithms have been recently used for the design of cloaking devices and lenses \cite{IserniaCloaking2019,IserniaScientificReports2017,IserniaGRIN2018}. Similarly, inverse source algorithms have also been suggested for design applications~\cite{BucciAntennaSynthesis}, and have been used, for example, in the design of shaped beam reflectarrays~\cite{MassaReflectarrayInverseSource}. 

Inverse scattering and inverse source problems are inherently ill-posed. That is, the solution to the associated mathematical problem has at least one of these features: (i)~non-uniqueness, (ii)~non-existence, or (iii)~instability~\cite{hansen1998rank}. The main reason behind non-uniqueness is non-radiating sources: current sources that may exist in the investigation domain but create no electromagnetic fields at the observation domain~\cite{DevaneyTAP1982}, i.e., creating a null-space for the mathematical problem. Although this non-uniqeness is typically a disadvantage in characterization, it is actually \textit{advantageous} for design as it provides extra degrees of freedom. For example, in \cite{MassaReflectarrayInverseSource}, the non-uniqueness due to non-radiating sources has been used to satisfy user-defined geometrical constraints in reflectarray design. On the other hand, non-existence of the solution does not occur in characterization applications, while it may occur in design applications, e.g., trying to design a highly directive radiation pattern from an electrically small aperture. In such situations, the electromagnetic inversion may still find a compromising solution that partially meets the desired design specifications. For example, consider a bianisotropic lens that transforms a given incident field to a desired field. The inverse scattering algorithm allows us to limit the solution space to purely isotropic dielectric lenses. In this situation, the solution to the mathematical problem may not exist; however, the inverse scattering algorithm may still result in an acceptable solution~\cite{IserniaGRIN2018}. Finally, the instability of the mathematical problem is related to the so-called smoothing effect \cite{Hansen1992} of the free space Green's function operator, which acts as a low-pass filter for high spatial frequency information. Mathematically, this instability occurs due to the very small singular values of the operator, and is treated by the use of appropriate regularization techniques which attempt to stabilize the problem by removing the effect of small singular values in the inversion process~\cite{Hansen2010Book,mojabi2009overview}.

We have recently proposed and used the electromagnetic inverse source framework for the design of metasurfaces \cite{brown2018use,brownAPS2019}. In addition to applying the inverse source framework to metasurface design, these works extend it to more practical scenarios: those where the desired specifications are expressed in terms of some performance criteria (e.g., HPBW and null directions) rather than a completely-known desired field pattern. In this paper, we explore this approach in detail. To this end, we start with a problem statement, then briefly describe our methodology, which is followed by a section motivating the pursuit of this approach. We then present some fundamentals of metasurfaces, followed by presenting the inverse source framework for design in the operator notation. The details of our inversion algorithms are finally presented followed by a brief description of the simulation algorithm. We then present some illustrative examples, explain our current limitations, and finally present our conclusions.

\section{Problem Statement}
\begin{figure}[t!]
\centerline{\includegraphics[width=3.4in]{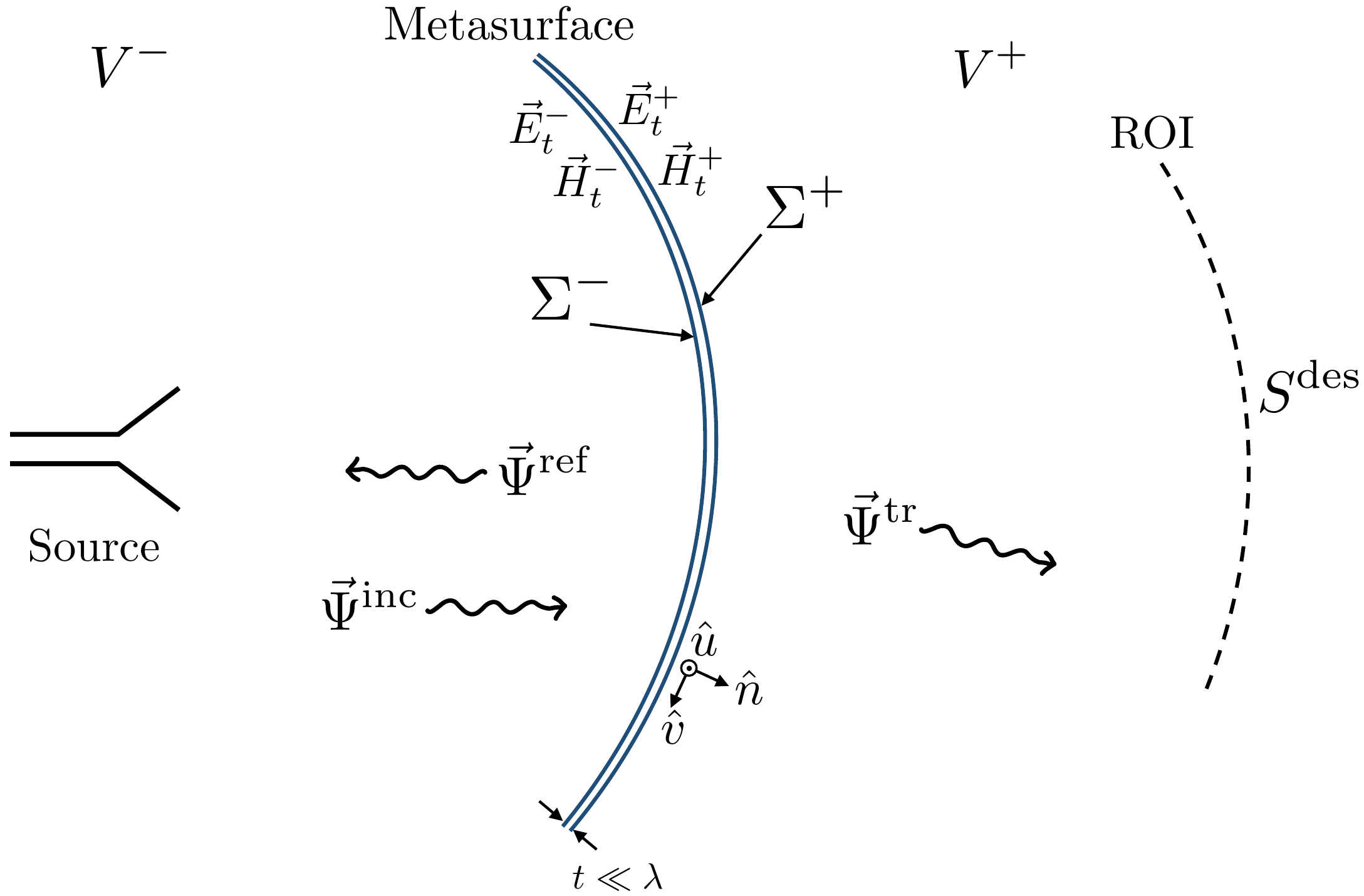}}
\caption{Input ($\Sigma^-$) and output ($\Sigma^+$) bounding surfaces of the metasurface on which the field transformation is specified. The variables $V^-$ and $V^+$ denote the input and output volumes bounded by the metasurface. The region of interest (ROI) is the region at which we would like to achieve some desired specifications. The incident field $\vec{\Psi}^\textrm{inc}$ generated by a source illuminates the input surface $\Sigma^-$ which then results in the reflected field $\vec{\Psi}^{\textrm{ref}}$ and the transmitted field $\vec{\Psi}^{\textrm{tr}}$. The tangential electric and magnetic fields on the input and output surfaces of the metasurface are denoted by $\vec{E}_t^-$, $\vec{H}_t^-$, $\vec{E}_t^+$, and $\vec{H}_t^+$. At any point on $\Sigma^+$, we define the local coordinate $(\hat{u},\hat{v},\hat{n})$ where $\hat{n}$ is perpendicular to $\Sigma^+$.}
\label{fig:MSSetup}
\end{figure}

\begin{figure}[t!]
\centerline{\includegraphics[width=3in]{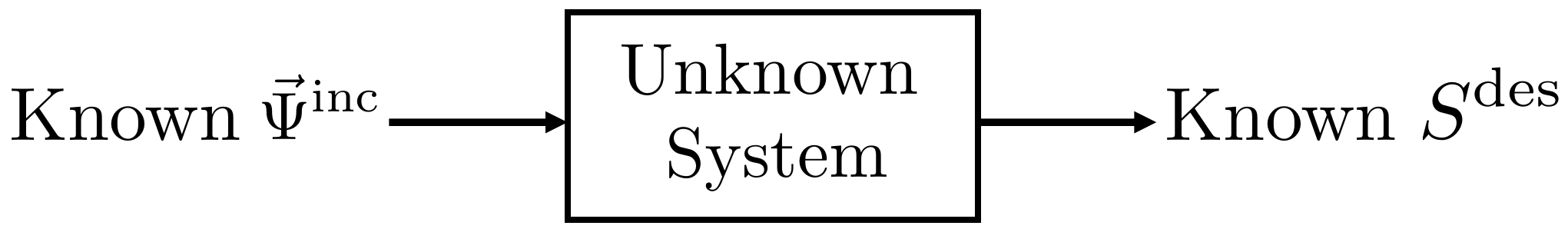}}
\caption{The design problem from a system viewpoint. The unknown properties of the system (surface susceptibilities of a metasurface) are to be found from the knowledge of the input and output. The known input to the system is the incident field $\vec{\Psi}^{\textrm{inc}}$ and the known output is the desired specifications $S^{\textrm{des}}$. Note that $S^{\textrm{des}}$ can be, for example, a set of performance criteria such as HPBW and null directions.}
\label{fig:system}
\end{figure}

Consider the geometry shown in Figure~\ref{fig:MSSetup}, where the surfaces $\Sigma^-$ and $\Sigma^+$ represent the input and output surfaces of the metasurface, and $V^-$ and $V^+$ represent the volumetric domains located on the input and output sides of the metasurface respectively.\footnote{To precisely define $V^-$ and $V^+$, we need to assume that $\Sigma^-$ and $\Sigma^+$ extend to infinity or are closed surfaces. However, the method presented herein relaxes this requirement.} Let us assume that the following items are known: ({i}) an incident field $\vec{\Psi}^{\textrm{inc}}$ where $\Psi \in \{E,H\}$ that impinges on the metasurface from the input side, ({ii}) a surface geometry $\Sigma$ of arbitrary shape for the metasurface, and ({iii}) a set of desired field specifications, denoted by ${S}^{\textrm{des}}$ in a region of interest (ROI) within the output side of the metasurface $V^+$. For example, the ROI can be a far-field region defined within a solid angle, or can be a near-field region. Our objective is to find the required electric and magnetic surface susceptibility profiles for the metasurface to transform $\vec{\Psi}^{\textrm{inc}}$ to a transmitted field, say $\vec{\Psi}^{\textrm{tr}}$, that exhibits the desired field specifications ${S}^{\textrm{des}}$. It should be noted that in contrast to existing design methods for metasurfaces, the proposed method does not assume the knowledge of $\vec{\Psi}^{\textrm{tr}}$; it only assumes the more practical knowledge of ${S}^{\textrm{des}}$, which could be for example a set of performance criteria such as HPBW or null directions. In summary, as depicted in Figure~\ref{fig:system}, we aim to infer the unknown properties of a system from a known input and its corresponding known output.

\section{Methodology}
\label{sec:method}
This section aims at providing a general idea about our methodology. (This topic will be re-visited with more details in Sections~\ref{sec:inversedesign} and \ref{sec:designframework}.) We start by noting that, as in existing design methods, e.g., see \cite{epstein2016huygens, achouri2015general}, the knowledge of the tangential electric and magnetic fields on the input  ($\Sigma^-$) and  output  ($\Sigma^+$) surfaces is needed to find the required surface susceptibility profiles of the metasurface. 

\subsection{Finding Tangential Fields on the Input Surface}
\label{subsec:internaltangentialfield}
The tangential electric and magnetic fields on ${\Sigma}^-$, denoted by $\vec{E}_t^-$ and $\vec{H}_t^-$, consist of two components: ({i})~the incident field on the metasurface $\vec{\Psi}^{\textrm{inc}}$, and ({ii})~the field $\vec{\Psi}^{\textrm{ref}}$ reflected from the metasurface. The incident field is often known since we illuminate the metasurface with a known antenna. In addition, the reflected field is assigned to be zero for reflectionless metasurfaces. Therefore, in this paper, we treat $\vec{E}_t^-$ and $\vec{H}_t^-$ as known quantities. 

\subsection{Finding Tangential Fields on the Output Surface}
\label{subsec:inversesource}
Let us denote the tangential electric and magnetic fields on $\Sigma^+$ by $\vec{E}_t^+$ and $\vec{H}_t^+$ respectively. The core of our methodology lies in inferring $\vec{E}_t^+$ and $\vec{H}_t^+$ from the knowledge of ${S}^{\textrm{des}}$. (Note that we do not assume the complete knowledge of $\vec{\Psi}^{\textrm{tr}}$; instead we merely assume the knowledge of $S^{\textrm{des}}$.) To this end, we use the electromagnetic inverse source framework to reconstruct required electric and magnetic currents ($\vec{J}$ and $\vec{M}$) on $\Sigma^+$ from the knowledge of ${S}^{\textrm{des}}$. Once these currents are reconstructed, we can then find $\vec{E}_t^+$ and $\vec{H}_t^+$.  Having obtained the tangential electric and magnetic fields on $\Sigma^-$ and $\Sigma^+$, we can find the required surface electric and magnetic susceptibility profiles for the metasurface based on the generalized sheet transition conditions (GSTCs)~\cite{idemen2011discontinuities, kuester2003averaged}. As will be discussed in Section~\ref{sec:limitations}, our current implementation does not enforce local power conservation or consider magnetoelectric coupling. Therefore, the resulting metasurfaces will include lossy and/or active elements.

\section{Motivation}
\label{sec:motivation}
Let us now motivate the utilization of the electromagnetic inverse source framework for finding $\vec{E}_t^+$ and $\vec{H}_t^+$. As will be seen, this has two main advantages: ({i})~the ability to work with arbitrarily-shaped geometries (in our case, various $\Sigma$ and ROI geometries), and ({ii})~the ability to handle various forms of ${S}^{\textrm{des}}$. To understand this, consider the following five cases.

\subsection{Case I}
If the desired output specification ${S}^{\textrm{des}}$ is a refracted plane wave in $V^+$, finding $\vec{E}_t^+$ and $\vec{H}_t^+$ is trivial as the expression for $\vec{\Psi}^{\textrm{tr}}$ is analytically known. (This is the case that has been mainly studied in the literature.) For this case, the electromagnetic inverse source framework is {not} needed.

\subsection{Case II}
Assume that the desired output specification ${S}^{\textrm{des}}$ is given in terms of the desired complex (amplitude and phase) field on a canonical surface (ROI) within $V^{+}$, and that the metasurface geometry ($\Sigma$) is of canonical shape. In such a case, $\vec{E}_t^+$ and $\vec{H}_t^+$ can be easily found by standard modal expansion algorithms. Such algorithms~\cite{parini2014theory} are widely used in planar, cylindrical, and spherical near-field antenna measurements to back-propagate the measured data to the surface of the antenna under test for diagnostics. These algorithms can also be used to back-propagate the desired complex field data from the ROI to $\Sigma^+$ if both the ROI and $\Sigma^+$ are canonical shapes. Case~II is in fact similar to Case~I, and this can be justified based on the plane wave spectrum as follows. Assume that both $\Sigma$ and ROI are parallel planar surfaces; then the modal expansion algorithm expands the desired complex field on the ROI in terms of a sum of plane waves (plane wave spectrum). We can then back-propagate all of these plane waves to the surface of the metasurface. Therefore, the use of electromagnetic inverse source algorithms is also not needed in Case~II.\footnote{It should be noted that since in practical applications we are dealing with truncated surfaces (e.g., finite planes), the use of electromagnetic inverse source framework can still offer an advantage compared to the modal expansion algorithms (which assume infinite planes or closed surfaces). This will not be further discussed in this paper; see \cite{petre1996differences} for a discussion on this topic as applied to near-field antenna measurements.} 

\subsection{Case III}
Assume that the desired output specification ${S}^{\textrm{des}}$ is given in terms of the complex field on an arbitrarily-shaped surface (ROI) within $V^+$ and/or the geometry of the metasurface ($\Sigma$) is of arbitrary shape~\cite{dehmollaian2019wave}. This could, for example, occur when we aim to design conformal metasurfaces. Standard modal expansion algorithms cannot handle such geometrical scenarios. In this case, the use of electromagnetic inverse source algorithms is advantageous. (This advantage of the electromagnetic inverse source algorithms has been utilized in near-field antenna measurements, e.g., see~\cite{ChaiANTEM2018}.)

\subsection{Case IV}
Assume that the desired specification ${S}^{\textrm{des}}$ is given in terms of amplitude-only (phaseless) fields. This could, for example, happen when the designer is interested in focusing the electromagnetic energy in a `hot spot' for microwave hyperthermia applications, or to design a desired power pattern for antenna applications. For this case, even if $\Sigma$ and the ROI are canonical surfaces, standard modal expansion algorithms are incapable of handling the phaseless nature of the data. One way to treat this is to assume a phase distribution; however, the phase assumption will automatically limit the number of achievable solutions. A better way is to avoid any phase assumptions and apply a phaseless electromagnetic inverse source algorithm to the data. We then typically obtain more solutions that satisfy the phaseless specifications, and single out the solution that results in the easiest physical implementation. (The use of phaseless electromagnetic inverse source algorithms has been previously considered in near-field antenna measurements to avoid the necessity of performing expensive phase measurements, e.g., see~\cite{brown2017multiplicatively}.)

\subsection{Case V}
Finally, case V aims at offering more flexibility to the designer. To this end, in Case~V, the desired specifications ${S}^{\textrm{des}}$ are some performance criteria rather than a field/power pattern. This is most practical for antenna engineering applications, e.g., setting the desired performance criteria to be the null directions or HPBW. This case cannot be handled by standard modal expansion algorithms, and we propose to use the electromagnetic inverse source framework to handle this case. 
\\~\\
In summary, there are at least three cases that existing metasurface synthesis methods cannot handle, and where the use of the electromagnetic inverse source framework is therefore advantageous.

 \section{Metasurface Fundamentals}
This section provides a brief overview of metasurface synthesis to show the importance of having the tangential field components on the input and output surfaces ($\Sigma^+$ and $\Sigma^-$) of the metasurface. The electromagnetic behaviour of a metasurface can be modelled by rigorous GSTCs, originally developed in~\cite{idemen2011discontinuities}. The form that we adopt herein was developed in~\cite{kuester2003averaged}, and we follow the notation style of~\cite{achouri2015general}\footnote{For other homogenization models of metasurfaces (polarizability-based model and equivalent impedance matrix model) see Section 2.4 in~\cite{yang2019surface}.}.

%
%


The electric and magnetic polarization densities can be expressed as
\begin{subequations}
\begin{align}
 \vec{P}^\text{e} &= \epsilon_0 \overline{\overline{\chi}}_{\text{ee}} \vec{E}_{\text{av}} + \overline{\overline{\chi}}_{\text{em}}\sqrt{\mu_0 \epsilon_0} \vec{H}_{\text{av}}, \label{eq:polarizationP}\\
 \vec{P}^\text{m} &= \overline{\overline{\chi}}_{\text{mm}} \vec{H}_{\text{av}} + \overline{\overline{\chi}}_{\text{me}}\sqrt{\frac{\epsilon_0}{\mu_0}} \vec{E}_{\text{av}},\label{eq:polarizationM}
\end{align}
\end{subequations}
where $\epsilon_0$ and $\mu_0$ are the permittivity and permeability of free space, respectively, and $\overline{\overline{\chi}}_{\text{ee}}$, $\overline{\overline{\chi}}_{\text{mm}}$, $\overline{\overline{\chi}}_{\text{em}}$, and $\overline{\overline{\chi}}_{\text{me}}$ represent the electric/magnetic (first subscript) surface susceptibility tensors describing the response to electric/magnetic (second subscript) field excitations~\cite{vahabzadeh2016simulation}. The average fields on the surface $\Sigma$ of the metasurface are defined as
\begin{equation}
 \vec{\Psi}_{\text{av}} \triangleq \frac{\vec{\Psi}^{\textrm{tr}}\vert_{\Sigma^+} + \left(\vec{\Psi}^{\textrm{inc}}\vert_{\Sigma^-} + \vec{\Psi}^{\textrm{ref}}\vert_{\Sigma^-}\right)}{2}.
\end{equation}

We now assume that the normal components of the polarization densities are zero (i.e., $\hat{n} \cdot \vec{P}^\text{e} = \hat{n} \cdot \vec{P}^\text{m} = 0$, where $\hat{n}$ is the outward unit vector normal to the metasurface as shown in Figure~\ref{fig:MSSetup}) to avoid the unnecessary complexity that would result from including spatial derivatives in the GSTCs~\cite{achouri2015general}. This assumption does not result in any loss of generality, as the normal components of any field can be represented in terms of the corresponding tangential components (i.e., allowing non-zero normal components of the polarization densities would not introduce any independent solutions). Assuming a time-dependency of $e^{j\omega t}$, we can now construct a system of equations from the GSTCs in~\cite{achouri2015general} and equations \eqref{eq:polarizationP} and~\eqref{eq:polarizationM}, relating the tangential electric and magnetic fields through the surface susceptibilities as
\begin{subequations}
\begin{align}
\begin{split}
 \begin{pmatrix}
 -\Delta H_v \\
 \Delta H_u
 \end{pmatrix}
 = &\textrm{ }j\omega \epsilon_0
 \begin{pmatrix}
 \Xee^{uu} & \Xee^{uv} \\ \Xee^{vu} & \Xee^{vv}
 \end{pmatrix}
 \begin{pmatrix}
 E_{u,\text{av}} \\  E_{v,\text{av}} 
 \end{pmatrix}
 \\ &\quad + j\omega \sqrt{\epsilon_0 \mu_0}
 \begin{pmatrix}
 \Xem^{uu} & \Xem^{uv} \\ \Xem^{vu} & \Xem^{vv}
 \end{pmatrix}
 \begin{pmatrix}
 H_{u,\text{av}} \\  H_{v,\text{av}}
 \end{pmatrix},
 \end{split}
 \\~ \nonumber \\
 \begin{split}
 \begin{pmatrix}
 -\Delta E_u \\
 \Delta E_v
 \end{pmatrix}
 = &\textrm{ }j\omega \mu_0
 \begin{pmatrix}
 \Xmm^{vv} & \Xmm^{vu} \\ \Xmm^{uv} & \Xmm^{uu}
 \end{pmatrix}
 \begin{pmatrix}
 H_{v,\text{av}} \\  H_{u,\text{av}} 
 \end{pmatrix}
 \\ &\quad + j\omega \sqrt{\epsilon_0 \mu_0}
 \begin{pmatrix}
 \Xme^{vv} & \Xme^{vu} \\ \Xme^{uv} & \Xme^{uu}
 \end{pmatrix}
 \begin{pmatrix}
 E_{v,\text{av}} \\  E_{u,\text{av}}
 \end{pmatrix},
\end{split}
\end{align}\label{eq:chifull}
\end{subequations}

\noindent where the subscripts and superscripts $u$ and $v$ refer to the tangential components of the local coordinate system defined by $\hat{u} \times \hat{v} = \hat{n}$ and $\hat{u} \perp \hat{v}$. The $\Delta$ operator denotes the difference between the fields on either side of the metasurface, defined in terms of the transmitted, incident, and reflected fields as
\begin{equation}
\Delta \vec{\Psi} \triangleq \vec{\Psi}^{\textrm{tr}} - \left(\vec{\Psi}^{\textrm{inc}} + \vec{\Psi}^{\textrm{ref}}\right).
\end{equation} 

For a single field transformation (i.e., one combination of incident, reflected, and transmitted fields), the system in~\eqref{eq:chifull} is clearly underdetermined. 
It should also be noted that solving~\eqref{eq:chifull} for a desired field transformation does not guarantee that the resulting susceptibilities are physically practical or realizable. Avoiding these impractical solutions requires enforcing additional constraints and/or modifying the generated solution. Finally, we note that all the field quantities in (\ref{eq:chifull}a) and (\ref{eq:chifull}b) can be found based on the knowledge of $\vec{E}_t^+$, $\vec{H}_t^+$, $\vec{E}_t^-$, and $\vec{H}_t^-$. For example,
\begin{equation}
\Delta E_v = \hat{v} \cdot (\vec{E}_t^+ - \vec{E}_t^-),
\end{equation}
\begin{equation}
H_{u,\text{av}} = \hat{u} \cdot \left(\frac{\vec{H}_t^+ + \vec{H}_t^-}{2}\right).
\end{equation}
Therefore, as noted in Section~\ref{sec:method}, having the knowledge of $\vec{E}_t^+$, $\vec{H}_t^+$, $\vec{E}_t^-$, and $\vec{H}_t^-$ enables us to find the required susceptibility profiles.  

\section{Inverse Source Framework}
\label{sec:inversedesign}
Herein, we extend Section~\ref{sec:method} by detailing how the inverse source framework for metasurface design is formulated. This is explained in the following seven steps.

\subsection{Preparing the Data to be Inverted}
\label{subsec:datapreparation}
The first step in the development of the inverse source design formulation is to ensure that the desired specification ${S}^{\textrm{des}}$ is in a proper form for use by inverse source algorithms. We consider the following three scenarios. (i)~If ${S}^{\textrm{des}}$ is in the form of complex fields, we directly use it  in the inverse source formulation. (ii)~If ${S}^{\textrm{des}}$ is in the form of power patterns (phaseless), we also directly use it in the formulation. (iii)~If ${S}^{\textrm{des}}$ is in the form of some performance criteria (e.g., main beam direction, HPBW and null directions), we first form a weighted normalized power pattern (phaseless) that satisfies the required performance criteria. We will then use this pattern in the inverse source algorithm. (This step will be covered in more details in Section~\ref{subsec:situation3}.) Therefore, in summary, the data to be used in the inversion algorithm can be represented in the form of $\mathcal{K}(S^{\textrm{des}})$, where $\mathcal{K}$ is the operator that converts $S^{\textrm{des}}$ to a data set that can be used by the inversion algorithm. 

\subsection{Creating the Data Operator}
The so-called data operator, denoted by $\mathcal{G}$, takes a given set of electric and magnetic currents ($\vec{J}$ and $\vec{M}$) on $\Sigma^+$ and outputs the corresponding complex fields on the ROI, i.e., $\vec{\Psi}^{\textrm{tr}}\rvert_{\textrm{ROI}} = \mathcal{G}(\vec{J},\vec{M})$. In our case, this operator is constructed based on the electric field integral equation (EFIE)~\cite{alvarez2007reconstruction}. In the three-dimensional (3D) case, the operator $\mathcal{G}$ can be expressed as \cite{quijano2009improved}
\begin{align}
\mathcal{G}(\vec{J},\vec{M}) = &-j\eta_0 k_0 \int \limits_{\Sigma^+} \left[ \vec{J}(\vec{r}^{\,\prime}) + \frac{1}{k_0^2}\nabla \nabla_s^{\prime} \cdot \vec{J}(\vec{r}^{\,\prime}) \right] g(\vec{r},\vec{r}^{\,\prime})ds^{\prime} \nonumber \\ &-\nabla \times \int \limits_{\Sigma^+} \vec{M}(\vec{r}^{\,\prime})  g(\vec{r},\vec{r}^{\,\prime}) ds^{\prime}
\end{align}
where the position vectors $\vec{r}$ and $\vec{r}^{\,\prime}$ belong to the ROI and $\Sigma^{+}$ respectively. In addition, $g(.,.)$ denotes the Green's function, and `$\nabla^{\prime}_s \cdot$' is the surface divergence operator with respect to the prime coordinate (i.e., position vector on $\Sigma^{+}$). Finally, $\eta_0$ and $k_0$ are the wave impedance and wavenumber in free space.\footnote{It should be noted that the use of Green's function formulation enables us to perform the design in other media which are not free space, assuming that the Green's function in those media are known either analytically or numerically.} 

\subsection{Forming the Data Misfit Cost Functional}
For various sets of $\vec{J}$ and $\vec{M}$ on $\Sigma^+$, we need to compare the simulated effects with the desired effect to converge at an appropriate set of equivalent currents. In other words, we need to  compare $\mathcal{G}(\vec{J},\vec{M})$ with $\mathcal{K}(S^{\textrm{des}})$ to evaluate how appropriate a set of $(\vec{J},\vec{M})$ is. Before doing so, we first need to make sure that $\mathcal{G}(\vec{J},\vec{M})$ is in the form that is consistent with $\mathcal{K}(S^{\textrm{des}})$. For example, for the power pattern synthesis problem, we need to compare $|\mathcal{G}(\vec{J},\vec{M})|^2$ with $\mathcal{K}(S^{\textrm{des}})$. To this end, let us assume that the operator $\mathcal{L}$ takes $\mathcal{G}(\vec{J},\vec{M})$ and outputs the form that is consistent with $\mathcal{K}(S^{\textrm{des}})$. We then form a data misfit cost functional that is the $L_2$ norm discrepancy between the desired effect $\mathcal{K}(S^{\textrm{des}})$ and the simulated effect $\mathcal{L}(\mathcal{G}(\vec{J},\vec{M}))$ for a given electric and magnetic currents. The data misfit cost functional $\mathcal{C}$ is then 
\begin{equation}
\mathcal{C}(\vec{J},\vec{M}) =  \norm{\mathcal{L}(\mathcal{G}(\vec{J}, \vec{M})) - \mathcal{K}({S}^{\textrm{des}})}^2_2,  
\label{eq:generalFunctional}
\end{equation}
where $\norm{\textrm{.}}_2$ denotes the $L_2$ norm which is defined over the ROI at which the desired specifications are defined.\footnote{As will be seen in Section~\ref{subsec:situation3}, we may add an extra cost functional to the data misfit cost functional to enforce additional specifications.}

\subsection{Enforcing Love's Equivalence Condition}
\label{subsec:love}
Based on the electromagnetic equivalence principle~\cite{rengarajan2000field}, when solving for $\vec{J}$ and $\vec{M}$ on $\Sigma^+$ from the knowledge of $S^\textrm{des}$ we have the following two general options regarding the fields internal to $\Sigma^{+}$: (i)~no assumptions, or (ii) assuming a particular field such as a null field (Love's equivalence condition). It was shown in \cite{quijano2009improved} for antenna diagnostics applications that the inverse source problem can result in different sets of $\vec{J}$ and $\vec{M}$ solutions based on different assumptions regarding the inner side of the reconstruction surface. However, they generate similar external fields on the ROI (hence the non-uniqeness of the inverse problem). Herein, we apply Love's equivalence condition when solving for $\vec{J}$ and $\vec{M}$. To enforce this condition, we create a virtual surface that is inwardly offset with respect to $\Sigma^+$, and then enforce zero tangential electric fields on this surface when solving for $\vec{J}$ and $\vec{M}$. (For details, see~\cite{quijano2009improved}.)

\subsection{Minimizing the Data Misfit Cost Functional}
\label{subsec:minimization}
To reconstruct an appropriate set of equivalent electric and magnetic currents, we minimize the data misfit cost functional, i.e., 
\begin{equation}
\textrm{appropriate }  (\vec{J},\vec{M}) = \underset{\vec{J},\vec{M}}{\mathrm{argmin}} \left \{ \mathcal{C}(\vec{J},\vec{M}) \right \}.
\label{eq:generalFunctional2}
\end{equation}
To this end, we use the conjugate gradient (CG) algorithm~\cite{CG}. In this algorithm, $\vec{J}$ and $\vec{M}$ are iteratively updated as
\begin{equation}
(\vec{J}_{k+1},\vec{M}_{k+1}) = (\vec{J}_{k},\vec{M}_{k}) + \beta_k \mathbf{v}_k
\end{equation}
where $\beta_k \in \mathbb{R}$ is the step length and $\mathbf{v}_k$ is the conjugate gradient direction. To find $\mathbf{v}_k$, we need to know the gradient of the cost functional with respect to the currents at the $k^{\text{th}}$ iteration which is denoted by $\mathbf{g}_k$.
\subsection{Regularization}
\label{subsec:reg}
When minimizing (\ref{eq:generalFunctional2}), we need to ensure that the instability of the ill-posed problem is properly handled. The methods used to treat this instability are often referred to as regularization methods~\cite{hansen1998rank,hansen1994regularization,Hansen2010Book}, and the methods that control the regularization weight are often called regularization parameter choice methods~\cite{HansenLcurve}. One common method in treating this instability is to augment the data misfit cost functional by a penalty cost functional, e.g., by  the $L_2$ norm of $\vec{J}$ and $\vec{M}$. This method is referred to as Tikhonov regularization, which has been widely used in various applications. Another regularization method is to use an iterative algorithm such as the conjugate gradient least squares method and limit (truncate) the number of iterations~\cite{HansenIterative},\cite{mojabi2009overview}. Herein, we use this truncated CG approach, although the truncation point is chosen in an \textit{ad hoc} way in the current implementation.

\subsection{Finding the Susceptibility Profiles}
\label{subsec:findingsusceptibility}
Having found the equivalent currents, and noting that Love's equivalence condition has been applied in solving the inverse source problem, we can easily obtain the required tangential fields on $\Sigma^+$ from the reconstructed currents as 
\begin{equation}
\vec{H}_t^+ = \vec{J} \quad \quad \textrm{ and } \quad \quad \vec{E}_t^+ = - \vec{M} .
 \label{eq:bclove}
 \end{equation}
 As noted in Section~\ref{subsec:internaltangentialfield}, we assume that we already know $\vec{E}_t^-$ and $\vec{H}_t^-$. Now that $\vec{E}_t^+$, $\vec{H}_t^+$, $\vec{E}_t^-$, and $\vec{H}_t^-$ are known, we can use (\ref{eq:chifull}) to find the susceptibility profiles. This completes the description of our inverse source framework for design.

\section{Inversion Algorithm Implementation}
\label{sec:designframework}
The previous section outlined the main steps for the inverse source framework for metasurface design. Herein, we explain how the inversion algorithm is implemented in the discrete domain. To simplify the discussion, consider the following specific three scenarios, where $\xbold$ denotes the (unknown) concatenated vector of equivalent electric and magnetic currents at the discretized surface $\Sigma^+$. As described in Section~\ref{subsec:findingsusceptibility}, once $\xbold$ is reconstructed, the required susceptibilities can be found.

\subsection{Scenario I}
For this scenario, assume that the desired specifications $S^{\textrm{des}}$ are a set of complex field values which can be either in the near-field or far-field zones. That is, $S^{\textrm{des}} = \fbold$ where $\fbold$ is a vector of complex electric (or, magnetic field) values representing the desired field at the ROI. Then, the cost functional (in the discrete domain) is simply
\begin{equation}
\mathcal{C}\left(\xbold\right) = \left\| \A \xbold - \fbold \right\|^2_2
\label{eq:complete}
\end{equation}
where $\A$ is the discretized integral operator that when operating on a given $\xbold$ produces the fields at the locations of interest on the ROI. (We note that this formulation is not restricted to canonical ROI and $\Sigma^+$.)  To enforce Love's condition (see Section~\ref{subsec:love}), we augment $\fbold$ by a number of zeros as $\fbold = \left[\fbold; \boldsymbol{0} \right]$, where the operator ``$\textrm{};\textrm{}$'' denotes vector concatenation and $\boldsymbol{0}$ denotes a column vector of zeros whose length is the same as the number of points at which we enforce null fields at the virtual Love's surface. Once $\fbold$ is augmented, the matrix $\A$ needs to be augmented accordingly. To reconstruct $\xbold$, minimization of~\eqref{eq:complete} over $\xbold$ is performed using the CG method, which is identical to CG minimization commonly used in the source reconstruction method for near-field antenna measurements~\cite{alvarez2007reconstruction}.
\subsection{Scenario II}
In this scenario, a specific far-field power  pattern (phaseless quantity) is desired. That is, $S^{\textrm{des}} = \left| \fbold\right|^2$ where $\left| \fbold\right|^2$ is a vector of the desired squared field amplitude (i.e., power) at the ROI. Then, the cost functional becomes
\begin{equation}
\mathcal{C}\left(\xbold\right) = \left\| \left|\A \xbold\right|^2 -\left| \fbold\right|^2 \right\|^2_2
\label{eq:phaseless}
\end{equation}
where  $\left| \A \xbold \right|^2$ denotes the magnitude-squared of the simulated field due to a predicted $\xbold$. In contrast to (\ref{eq:complete}), the cost functional (\ref{eq:phaseless}) is nonlinear due to the absence of phase information.\footnote{It is instructive to note that when inverse source algorithms are used in phaseless near-field antenna measurement applications, the phaseless data $\left| \fbold\right|^2$ often need to be measured in two different regions so that the phase information can be implicitly derived using the correlation of the two phaseless data sets. However, in design applications, this is not necessary since we do not need to worry about the true phase data. In fact, we are only interested in {a} solution that satisfies the amplitude-only requirement.} This cost functional is minimized using the CG algorithm where the gradient vector at the $k^{\text{th}}$ iteration (evaluated at $\xbold_{k}$) can be derived as~\cite[Section 4.2.1]{brown2016antenna}
\begin{equation}
\mathbf{g}_k = 2\A^H \left[\mathbf{r}_{k} \odot \A \xbold_{k}\right].
\label{eq:phaselessgradient}
\end{equation}
The superscript $H$ denotes the Hermitian (complex conjugate transpose) operator, $\mathbf{r}_{k} = \left| \A \xbold_{k} \right|^2 - \left| \fbold \right|^2$ is the residual vector and $\odot$ represents a Hadamard (element-wise) product. Once the gradient vector $\mathbf{g}_k$ is found, the conjugate gradient direction $\mathbf{v}_k$ can be obtained. The step length $\beta_k$ at each CG iteration can also be found by minimizing $\mathcal{F}(\beta_k)=\mathcal{C}(\xbold_k + \beta_k \mathbf{v}_k)$ with respect to $\beta_k$~\cite[Section 4.2.2]{brown2016antenna}. This cost functional is a fourth-degree polynomial with respect to $\beta_k$ as
\begin{align}
\mathcal{F}(\beta_k) &= \beta_k^4  \left\| \left|\A \mathbf{v}_k\right|^2 \right\|^2_2 + 4\beta_k^3 \operatorname{Re}\langle \left|\A \mathbf{v}_k\right|^2, (\A \xbold_k) \odot (\A \mathbf{v}_k)^*\rangle \nonumber \\
&+ 2\beta_k^2 \left[ \operatorname{Re}\langle \mathbf{r}_k, \left|\A \mathbf{v}_k\right|^2 \rangle +  \left\| \operatorname{Re}((\A \xbold_k) \odot (\A \mathbf{v}_k)^*) \right\|^2_2 \right] \nonumber \\
&+ 4\beta_k \operatorname{Re}\langle \mathbf{r}_k, (\A \xbold_k) \odot (\A \mathbf{v}_k)^*\rangle +  \left\| \mathbf{r}_k\right\|^2_2
\label{eq:polynomial}
\end{align}
where $\operatorname{Re}$ is the real-part operator, $\langle .,.\rangle$ denotes the inner product operator, and the superscript $*$ denotes the complex conjugate operator. To find the step length $\beta_k$, we differentiate (\ref{eq:polynomial}) with respect to $\beta_k$, and then numerically solve for the roots of the resulting third-degree polynomial, which has a pair of complex conjugate roots and one real root. The step length will then be the real root. 
\subsection{Scenario III}
\label{subsec:situation3}
This scenario considers a more practical case in which the desired field is only known in terms of certain far-field performance criteria, such as main beam direction, HPBWs, and null directions. (We will consider polarization later in this section.) Thus, for example,
\begin{equation}
S^{\textrm{des}} = \{\textrm{main beam directions}, \textrm{HPBWs}, \textrm{nulls} \}.
\end{equation}
\subsubsection{Creating a Normalized Power Pattern from Performance Criteria} 
\begin{figure}[t!]
\centerline{\includegraphics[width=\columnwidth]{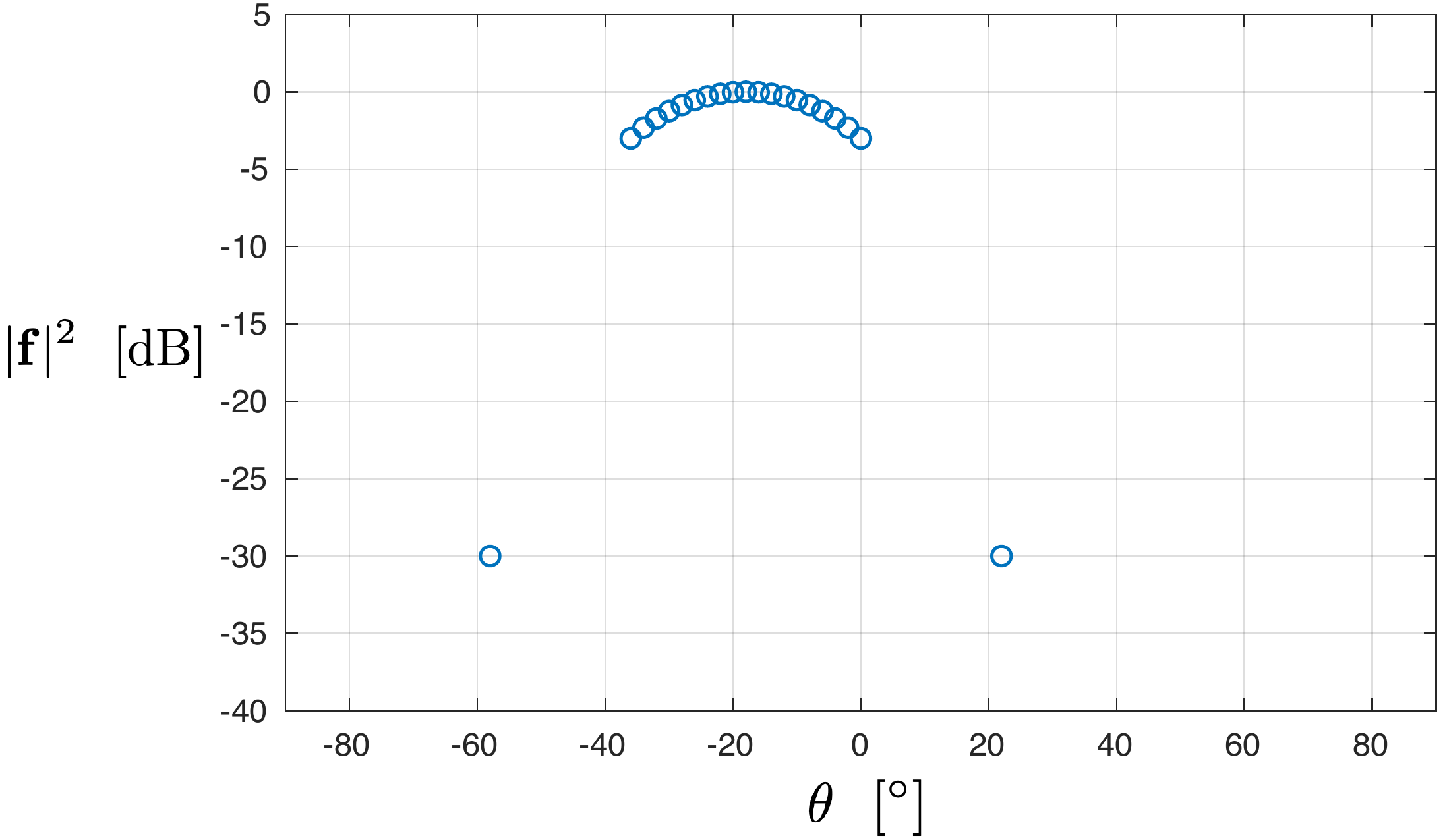}}
\caption{The normalized power pattern elements of $\left|\fbold\right|^2$ for a desired 2D far-field pattern with a main beam at $\theta=-18^{\circ}$, a HPBW of $36^{\circ}$, and a $-30$~dB null at $40^{\circ}$ away from the main beam. Note the two blue circles at $-30$~dB indicating the desired nulls.}
\label{fig:fboldExample1}
\end{figure}

The above desired performance criteria first need to be converted to a form that is usable by the inverse source algorithm, which is a normalized power pattern.\footnote{Normalization of the power pattern is not absolutely necessary, but allows for easier balancing if multiple cost functional terms are involved.} To this end, we form a power vector denoted by $\left| \fbold \right|^2$. In the direction of each main beam, the associated element of $\left| \fbold \right|^2$ is set to a value of 1, representing the normalized maximum power of the field. Nulls are represented in $\left|\fbold\right|^2$ by specifying the desired power level of the null relative to the main beam (e.g., a value of $10^{-3}$ would indicate a desired null at 30 dB below the main beam). The desired HPBWs for each main beam are enforced by approximating the main beam(s) using a cosine distribution. Desired power values are then interpolated using this approximation in a symmetric fashion around each main beam, resulting in corresponding elements in $\left|\fbold\right|^2$ ranging from 0.5 (at the HPBW limit) to 1 at the main beam. As an example, consider a desired far-field power pattern with a main beam at $\theta=-18^{\circ}$, a HPBW of $36^{\circ}$, and a $-30$~dB null at $40^{\circ}$ away from the main beam. For a 2D problem, this set of performance criteria would result in $\left|\fbold\right|^2$ taking the normalized power values shown in Figure~\ref{fig:fboldExample1}.

\subsubsection{Balancing the Normalized Power Pattern} 

As can be seen in Figure~\ref{fig:fboldExample1}, the quantitative values of the nulls can be significantly smaller than those of the main beams. In such situations, the inversion algorithm may favour main beam directions and HPBWs as compared to null directions.  To avoid this scenario, we balance $\left|\fbold\right|^2$ by the use of a diagonal `weighting' matrix denoted by $\mathbf{W}$. Now instead of inverting $\left|\fbold\right|^2$, we invert $\mathbf{W} \left|\fbold\right|^2$. The elements of $\mathbf{W}$ are selected to increase the relative contributions of the desired nulls to the functional, which can often be rendered insignificant compared to the main beam(s). More specifically, the $i$th diagonal element of $\mathbf{W}$, denoted by $w_i$, is computed as
\begin{equation}
\label{eq:weighting}
w_i = 
\begin{cases} 
      \frac{\zeta}{|f_i|^2}, & |f_i|^2 \leq  \gamma\\
     1, & \text{otherwise}
   \end{cases}
\end{equation}
where $\gamma$ is a real positive threshold parameter used to determine which data ($\left|f_i\right|^2$) is classified as a `null' and $\zeta$ is a real positive parameter that controls the amount of weighting applied, relative to the main beam.\footnote{This balancing idea has also been used in near-field antenna measurements~\cite{brown2018CAMA} and microwave imaging applications~\cite{mojabiPrescaled}.} 

\subsubsection{Creating a Data Misfit Cost Functional}

The data misfit cost functional is formed as
\begin{equation}
\mathcal{C}_1\left(\xbold\right)  = \designfxnaloneRHS .
\label{eq:specC1}
\end{equation}
Note that since these performance criteria are far-field quantities, $\left|\mathbf{A}\mathbf x\right|^2$ has been written as the summation of two terms: $\left|\mathbf{A}_{\xi1}\mathbf x\right|^2$ and $\left|\mathbf{A}_{\xi2}\mathbf x\right|^2$ where the subscripts $\xi_1$ and $\xi_2$ denote two \textit{locally} orthogonal polarizations of the field. 

\subsubsection{Enforcing the Polarization Specification} Assume the desired specification $S^{\textrm{des}}$ also includes a polarization constraint. This can be handled by adding the following cost functional to (\ref{eq:specC1})
\begin{equation}
\designfxnaltwo
\label{eq:xpol}
\end{equation}
where $\A_{\text{X-pol}}$ is the discrete integral operator related to the undesired polarization (the cross-polarized component) of the field. Therefore, the total cost functional may now be represented as
\begin{equation}
\mathcal{C}\left(\xbold\right) = \mathcal{C}_1\left(\xbold\right)  + \tau \mathcal{C}_2\left(\xbold\right)
\label{eq:specfunctional}
\end{equation}
where $\tau$ is a positive real parameter that controls the relative weight of the two cost functionals.

\subsubsection{Minimization}

To minimize~\eqref{eq:specfunctional} using the CG algorithm, the gradient vector at the $k^{\text{th}}$ iteration can be obtained as
\begin{align}
&\mathbf{g}_k= 2 {\mathbf{A}_{\xi 1}^H \left(\mathbf{W}^H \mathbf{r}_{1,k} \odot \mathbf{A}_{\xi 1}\mathbf{x}_{k} \right) + 2\mathbf{A}_{\xi 2}^H \left(\mathbf{W}^H \mathbf{r}_{1,k} \odot \mathbf{A}_{\xi 2}\mathbf{x}_{k} \right)} \nonumber \\
& \qquad \qquad+ 2\tau\mathbf{A}_{\text{X-pol}}^{H} \mathbf{A}_{\text{X-pol}} \mathbf{x}_k
\label{eq:specgradients}
\end{align}
where $\mathbf{r}_{1,k}$ is equal to the quantity inside the norm of \eqref{eq:specC1} evaluated at $\mathbf{x}_k$. A closed-form expression for the step length is not available, so instead an appropriate step length is found numerically using a technique known as Armijo's condition~\cite{dai2002conjugate}. It should be noted that there are several tuning parameters involved in Armijo's condition that affect the convergence of the minimization process. The stability of the iterative method can be very sensitive to these parameters, and finding appropriate values can often be a challenge.
Another difference when solving~\eqref{eq:specfunctional} compared with the two previous scenarios is how Love's condition is applied. The functional in~\eqref{eq:specfunctional} is first minimized \textit{without} Love's condition to generate an initial guess for the equivalent currents that is then used to solve~\eqref{eq:specfunctional} again \textit{with} Love's condition. We have found this two step procedure to greatly improve the convergence and stability of the solution. Lastly, it should be noted that the functionals in~\eqref{eq:specC1} and~\eqref{eq:xpol} are for the general case with fields radiating in 3D. For 2D problems, $\A_{\xi2}$ is simply set to zero and the same formulation can be used as written. In a similar fashion, $\mathbf{A}_{\text{X-pol}}$ is set to zero if a desired polarization is not required.

\section{FDFD-GSTC Solver}
Once we have reconstructed $\mathbf{x}$ and have subsequently obtained the required surface susceptibilities, we need to verify that they can, in fact, transform the incident field to a transmitted field that satisfies the required specifications $S^{\textrm{des}}$. To do this for 2D examples, we use a finite difference frequency domain (FDFD) GSTC solver, which is referred to as the FDFD-GSTC solver. This solver is an extension of the standard FDFD formulation with added boundary conditions to enforce the GSTC relationships. For details of the FDFD-GSTC solver, see \cite{vahabzadeh2016simulation}. 

It should be noted that the FDFD-GSTC solver is currently unable to simulate active metasurface elements (i.e., susceptibilities that correspond to gain). To overcome this limitation, we scale the  $E_t^+$ and $H_t^+$ achieved in (\ref{eq:bclove}), which is justified by noting that any scaled forms of (\ref{eq:bclove}) still satisfy $S^\textrm{des}$. Therefore, in our implementation, we scale both $\vec{H}_t^+$ and $\vec{E}_t^+$ by the coefficient $\alpha$ (i.e., $\vec{H}_t^+ = \alpha \vec{J}$ and $\vec{E}_t^+ = - \alpha \vec{M} $) to ensure that no active elements are required (resulting in increased loss). Finally, we note that since our current implementation of the FDFD-GSTC solver is limited to 2D problems, for 3D problems, we use a 3D forward EFIE electromagnetic solver to verify our reconstruction results. However, as opposed to the FDFD-GSTC solver, our 3D~EFIE solver does not incorporate the surface susceptibilities into its formulation, and merely calculates the effects of the reconstructed currents.

\section{Illustrative Examples}

This section presents several 2D and 3D synthetic examples demonstrating the flexibility of the inverse source method in dealing with different forms of desired specifications. Although not a requirement, in the examples presented here we consider the design of reflectionless metasurfaces and thus enforce $\vec{\Psi}^{\text{ref}} = 0$.

\subsection{2D Examples}

We first consider the design of several metasurfaces to transform an incident TE$_z$ wave, with fields propagating in 2D in the $xy$ plane at 10~GHz. The metasurface is 1D in this case, placed along the the line $x=0$ with elements (unit cells) that are $\lambda/6$ in length for each of the examples. ($\lambda$ denotes the wavelength.) Since all fields are TE$_z$, the system in~\eqref{eq:chifull} can be simplified and the metasurface can be represented by only four unknown susceptibility terms. Furthermore, we stipulate that there is no magnetoelectric coupling for simplicity, resulting in $\overline{\overline{\chi}}_{\text{em}} = \overline{\overline{\chi}}_{\text{me}} = 0$~\cite{vahabzadeh2016simulation}. With this choice, the unknown susceptibility tensors reduce to
\begin{subequations}
\begin{eqnarray}
 \Xee^{yy} = \frac{2}{j\omega \mu_0}\frac{E_y^{\text{inc}} + E_y^{\text{ref}} -E_y^{\text{tr}}}{H_z^{\text{inc}} + H_z^{\text{ref}} + H_z^{\text{tr}}} \label{eq:chi2D1}\bigg \rvert_{\textrm{on the metasurface}},\\
 \Xmm^{zz} = \frac{2}{j\omega \epsilon_0}\frac{H_z^{\text{inc}} + H_z^{\text{ref}} - H_z^{\text{tr}}}{E_y^{\text{inc}} + E_y^{\text{ref}} +E_y^{\text{tr}}}\bigg \rvert_{\textrm{on the metasurface}}.
\label{eq:chi2D2}
\end{eqnarray}
\end{subequations}
Note that in the above equations, all of the field components are evaluated tangential to the surface of the metasurface (i.e., $x=0$). Also, as explained earlier, $E_y^{\text{tr}}$ and $H_z^{\text{tr}}$ are not necessarily known on the surface of the metasurface, thus, justifying the use of the inverse source framework. In each example, the tangential components of the desired transmitted field (i.e., $E_y^{\text{tr}}$ and $H_z^{\text{tr}}$) are found using the inverse source step of the design procedure, and then~\eqref{eq:chi2D1} and \eqref{eq:chi2D2} are used to compute the required $\Xee^{yy}$ and $\Xmm^{zz}$ to support the transformation from a specified incident field with no reflections. The 2D examples are then verified using the FDFD-GSTC solver, which simulates the interaction between the incident field and the susceptibilities in a solution domain of $20\lambda$ by $30\lambda$ in the $xy$ plane.  The solution domain uses a discretization size of $\lambda/50$ and is bounded by a perfectly matched layer (PML) of thickness $\lambda$ on all sides. 

\subsubsection{Complex field}

\begin{figure}[!t]
\centerline{\includegraphics[width=\columnwidth]{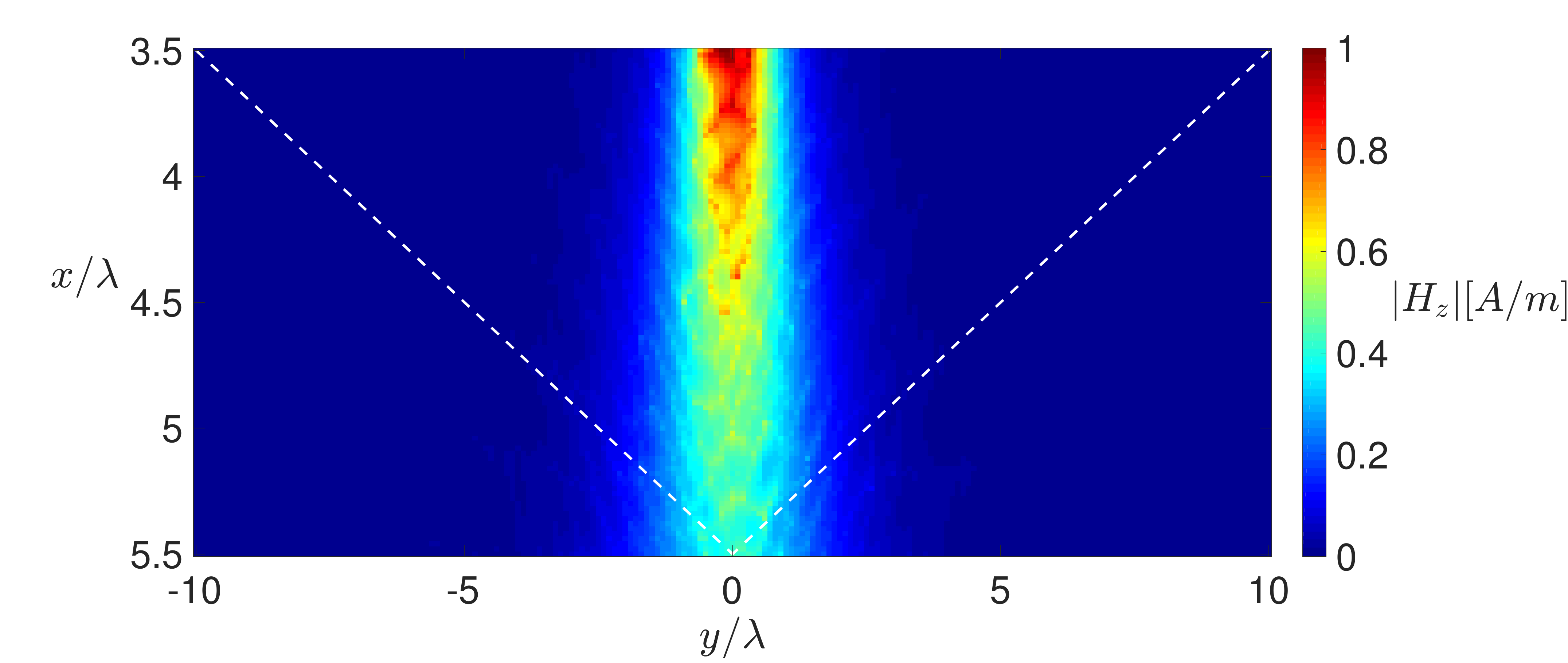}}
\caption{The desired NF distribution which is generated from a pyramidal horn in ANSYS HFSS. The amplitude and phase of $H_z$ are extracted along the dashed white lines (ROI) and used as input ($S^{\textrm{des}}$) for the metasurface design.}
\label{fig:desiredNF}
\end{figure}

\begin{figure}[!t]
\centerline{\includegraphics[width=\columnwidth]{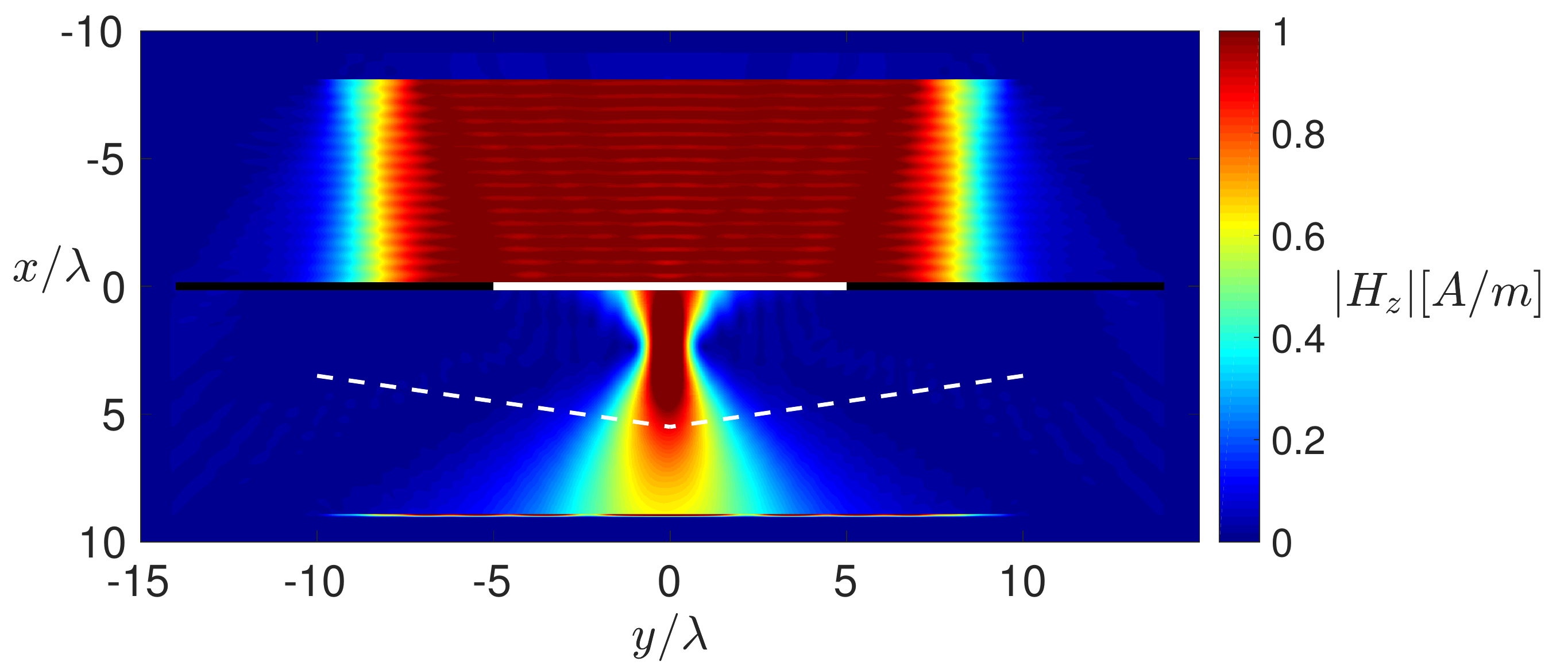}}
\caption{The magnitude of $H_z$ in the solution domain when the metasurface designed for the near-field distribution is illuminated by a normally incident tapered plane wave, simulated using the FDFD-GSTC solver. The designed metasurface extends from $y=-5\lambda$ to $y=5\lambda$ along $x=0$ as indicated by the solid white line, while absorbing susceptibilities have been added along $x=0$ for $|y| > 5\lambda$ as shown by the solid black line. The dashed white lines (ROI) indicate the location where the desired near-field $H_z$ is specified.}
\label{fig:completeNF_MS}
\end{figure}

\begin{figure}
         \centering
         \subfloat[]{\includegraphics[width=3.4in]{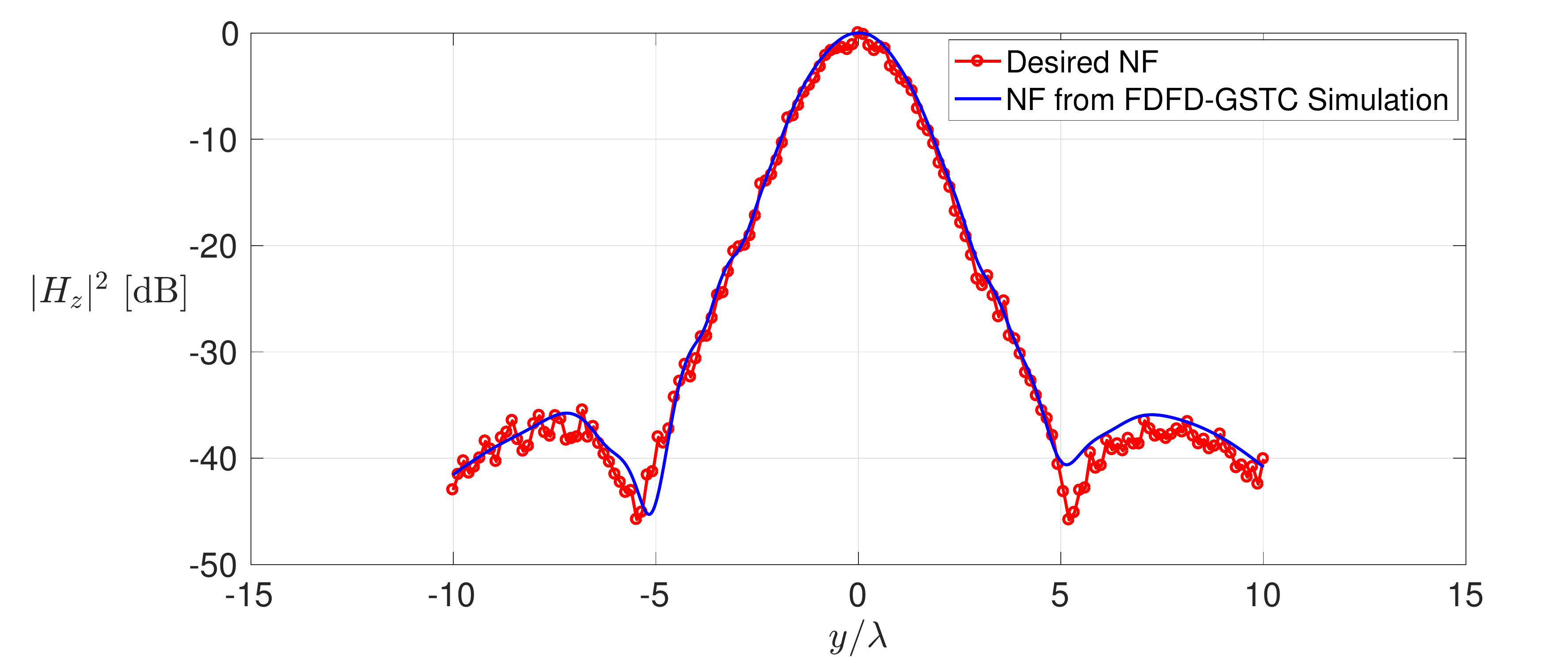}}\\
         \subfloat[]{\includegraphics[width=3.4in]{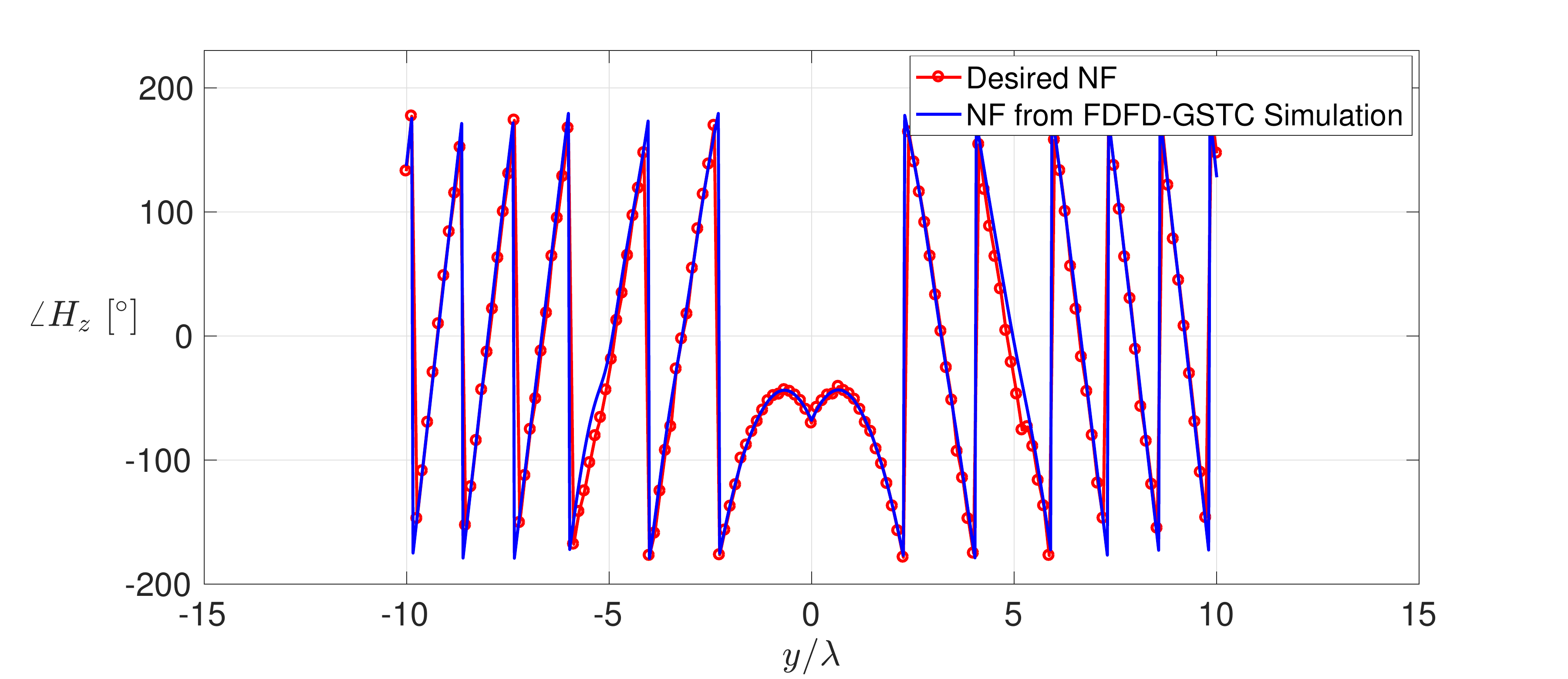}}
	\caption{A comparison of the desired normalized near-field (NF) distribution (solid red curve with circular markers) and the near-field distribution produced by the designed metasurface using both amplitude and phase information (solid blue curve).}
         \label{fig:completeComparison}
\end{figure}

For the first example, we attempt to produce a desired near-field (NF) distribution using complete (amplitude and phase) field information over a given ROI. The near-field distribution shown in Figure~\ref{fig:desiredNF} is generated from a pyramidal horn antenna in ANSYS HFSS at 10~GHz, and we use the complex field values ($H_z$) along the intersecting dashed white lines as input to the inverse source algorithm. That is, in this case, the intersecting dashed white lines are the ROI, and the complex near-field $H_z$ data on these white lines are  $S^{\textrm{des}}$.  Moreover, intentionally, the two white lines have been chosen to represent an ROI which is not of canonical shape. The designed portion of the metasurface extends from $y = -5\lambda$ to $y=5\lambda$ along the line $x=0$, with absorbing susceptibilities placed along the remainder of the $x=0$ line. The surface over which we reconstruct the equivalent currents coincides with the metasurface boundary $\Sigma$ and is discretized into elements of length $\lambda/6$. Love's equivalence condition is enforced along the line $x = -\lambda/10$ with a resolution of $\lambda/9$.
The proposed method is used to first find equivalent currents $J_y$ and $M_z$ on $\Sigma$ that produce the desired near-field by minimizing~\eqref{eq:complete}, and then~\eqref{eq:bclove} is used to calculate the desired tangential transmitted fields on the metasurface. The incident field in each of the 2D examples is a normally incident TE$_z$ plane wave, which is `tapered' along $y$ to avoid interactions with the PML that would introduce numerical error. (The $H_z$ amplitude of the plane wave is 1~$\textrm{A}/\textrm{m}$ for $|y| \leq 7\lambda$ and linearly decreases to zero for $7\lambda < |y| \leq 10\lambda$.) The necessary susceptibilities for the transformation (assuming no reflection) are then found using~\eqref{eq:chi2D1} and \eqref{eq:chi2D2}.
The total field produced by simulating the designed metasurface and the incident field with the FDFD-GSTC solver is shown in Figure~\ref{fig:completeNF_MS}. The efficiency, calculated as the ratio of power transmitted through the metasurface to the power incident on the metasurface, is 18.6\% for this example, thus, showing the presence of lossy regions within the metasurface.
Figure~\ref{fig:completeComparison} compares the desired (red) and produced (blue) amplitude and phase of $H_z$ along the dashed white lines, parametrized along $y$ only. As can be seen from the plot, both the desired amplitude and phase of $H_z$ have been produced with excellent accuracy.

\subsubsection{Phaseless Power Pattern}
\begin{figure}[!t]
\centerline{\includegraphics[width=\columnwidth]{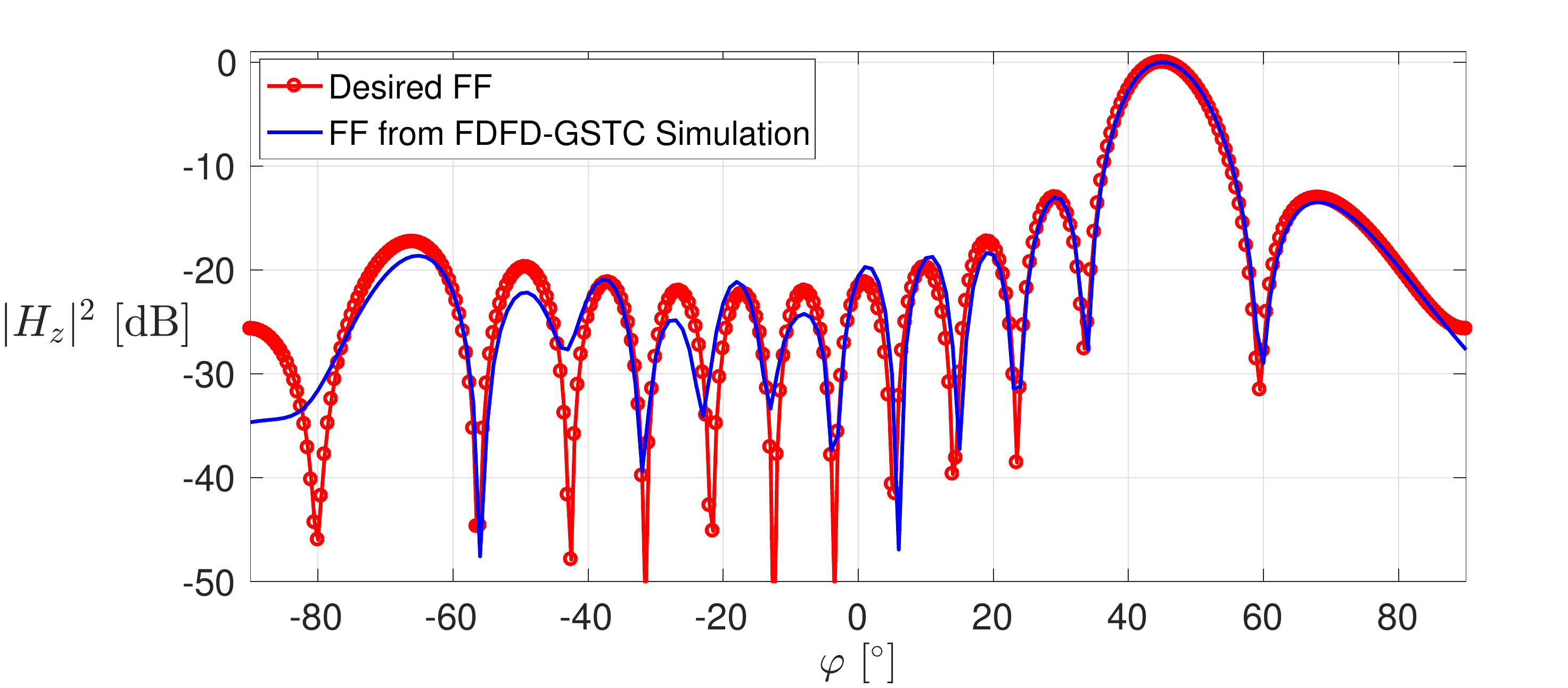}}
\caption{A comparison of the normalized far-field (FF) power pattern (phaseless) produced by the FDFD-GSTC simulation of the designed metasurface (solid blue curve) and the desired power pattern (solid red curve with circular markers).}
\label{fig:phaselessFF}
\end{figure}

\begin{figure}[!t]
\centerline{\includegraphics[width=\columnwidth]{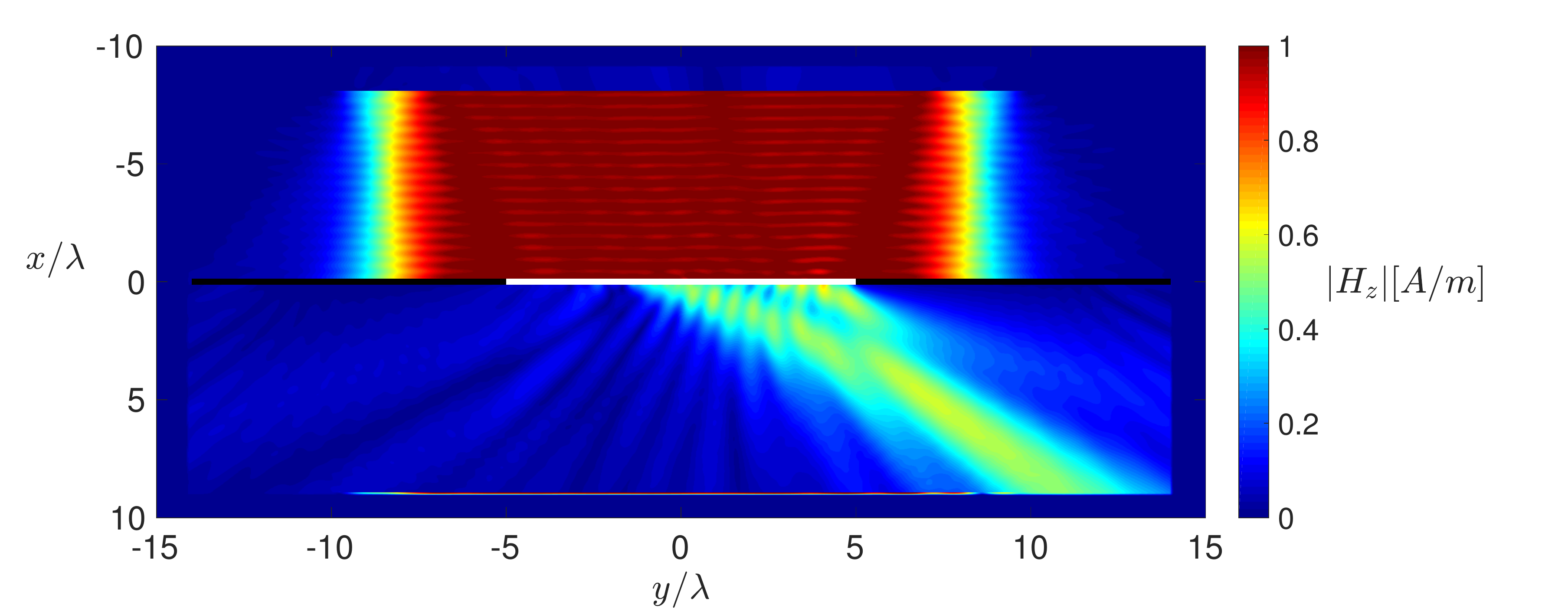}}
\caption{The magnitude of $H_z$ in the solution domain when the metasurface designed for the phaseless power pattern is illuminated by a normally incident tapered plane wave, simulated using the FDFD-GSTC solver.}
\label{fig:phaselessNF}
\end{figure}

The second example consists of attempting to produce a desired power pattern (phaseless) in the far-field (FF) region. In this example, the metasurface dimensions and incident field are the same as the previous example (relative to wavelength), with a frequency of operation of 1~GHz. The desired far-field power pattern for this example has been generated using a uniformly spaced array of 13 $\hat{z}$-directed elementary dipoles placed along the line $x=0$ from $y=-3\lambda$ to $y=3\lambda$. The squared amplitude (power) of $H_z$ produced by this array is computed on a semicircular domain of radius $500\lambda$ for $-90^{\circ} \leq \varphi \leq 90^{\circ}$ as shown in Figure~\ref{fig:phaselessFF} (solid red curve with circular markers). The required field components $E_y^{\text{tr}}$ and $H_z^{\text{tr}}$ (tangential to the metasurface) are found by minimizing~\eqref{eq:phaseless} and applying~\eqref{eq:bclove}, and then the required susceptibilities are found using (\ref{eq:chi2D1}) and (\ref{eq:chi2D2}). To evaluate the inversion performance, the resulting susceptibilities are simulated using the FDFD-GSTC solver. The total field in the solution domain is shown in Figure~\ref{fig:phaselessNF} to demonstrate the nearly reflectionless performance of the metasurface. The far-field power pattern produced by the incident field passing through the metasurface is shown in Figure~\ref{fig:phaselessFF} (solid blue curve), with an efficiency of 17.0\%. As can be seen from the plot, the produced far-field power pattern exhibits very good agreement with the desired power pattern.

\subsubsection{Performance Criteria}

\begin{figure}[!t]
\centerline{\includegraphics[width=\columnwidth]{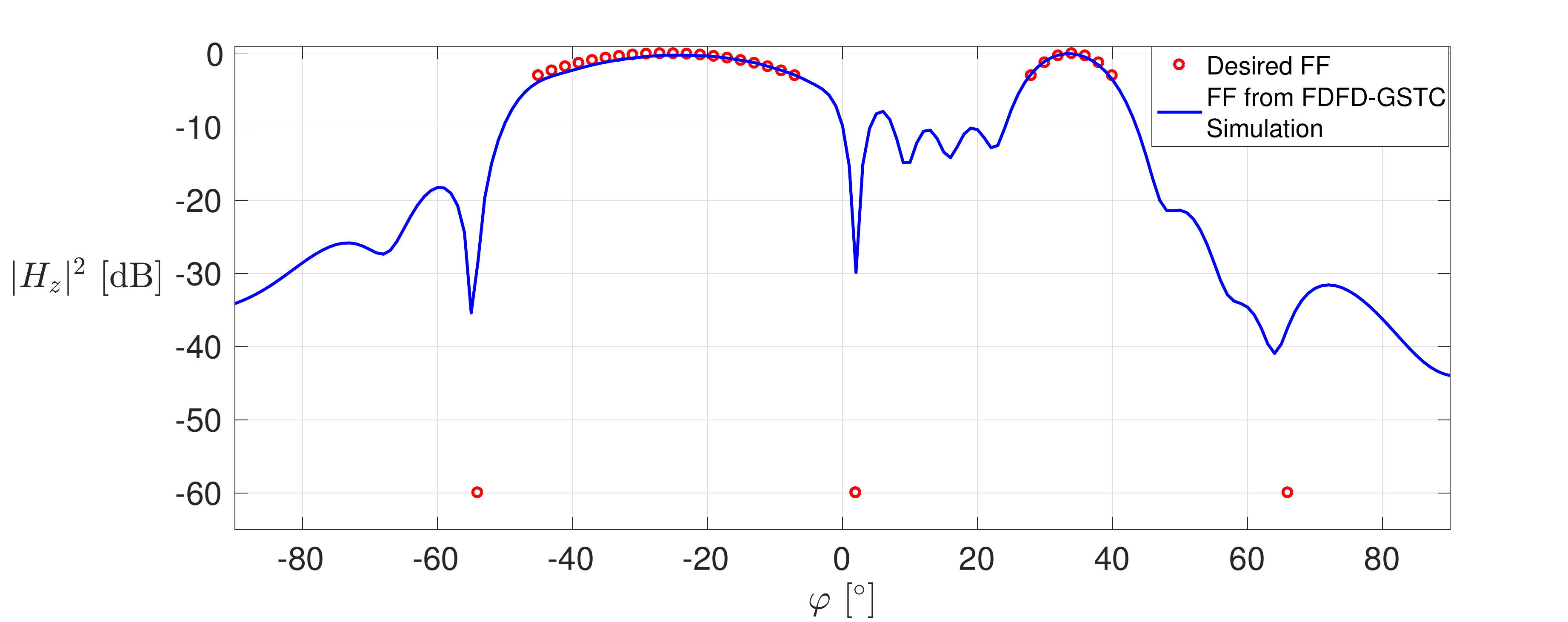}}
\caption{The red circular markers show the desired normalized far-field (FF) power pattern obtained from a set of performance criteria $S^{\textrm{des}}$ outlined in Table~\ref{tab:2Dspecs}. Note the three red circles at $-60$~dB level indicating the desired null directions. The solid blue curve shows the normalized FF power pattern produced by the FDFD-GSTC simulation of the designed metasurface.}
\label{fig:2DSpecsFF}
\end{figure}

The goal of the third example is to satisfy a set of desired far-field performance criteria. These specifications consist of two main beams with associated HPBWs and nulls as shown in Table~\ref{tab:2Dspecs}.
\begin{table}[t!]  
\renewcommand{\arraystretch}{1.3}
\caption{Desired Far-field (FF) Specifications -- 2D Example ($\theta=90^{\circ}$)}
\label{tab:2Dspecs}
  \centering
  \begin{tabular}{l | l | l}
   \textbf{Specifications} & \textbf{Main Beam 1} & \textbf{Main Beam 2} \\ \hline
Direction & $\varphi = -26^{\circ}$ & $\varphi = 34^{\circ}$\\ 
HPBW &$38^{\circ}$ & $12^{\circ}$ \\
{Nulls (relative to the main beam)} &  $28^{\circ}$ at $-60$~dB & $32^{\circ}$ at $-60$~dB\\
\end{tabular}
\end{table}
The metasurface and incident field in this case are the same as the previous example for the phaseless power pattern. The desired normalized power pattern created based on the desired performance criteria is shown in Figure~\ref{fig:2DSpecsFF} by red circular markers. In particular, note the three circular markers at $-60$~dB level indicating the desired null directions. The inversion algorithm finds the required equivalent currents by minimizing (\ref{eq:specC1}), and then the required tangential fields and subsequently the required susceptibilities are found. The resulting susceptibilities are then given to the FDFD-GSTC method to evaluate the performance of the metasurface. The far-field normalized power pattern produced by the metasurface designed for these specifications is shown in Figure~\ref{fig:2DSpecsFF} (blue curve) and represents an efficiency of 13.0\%. 
This achieved FF power pattern has main beams in  $\varphi = -25^{\circ}$ and $\varphi = 34^{\circ}$, with HPBWs of $36.1^{\circ}$ and $11.8^{\circ}$ respectively. The desired nulls, although not at the specified level of $-60$~dB, are clearly present in the FF power pattern as well. Comparing the metrics of the produced FF power pattern to those specified in Table~\ref{tab:2Dspecs} confirms that the resulting metasurface satisfies the design constraints with only a few minor differences.

\subsection{3D Example}
\begin{figure}[!t]
\centerline{\includegraphics[width=\columnwidth]{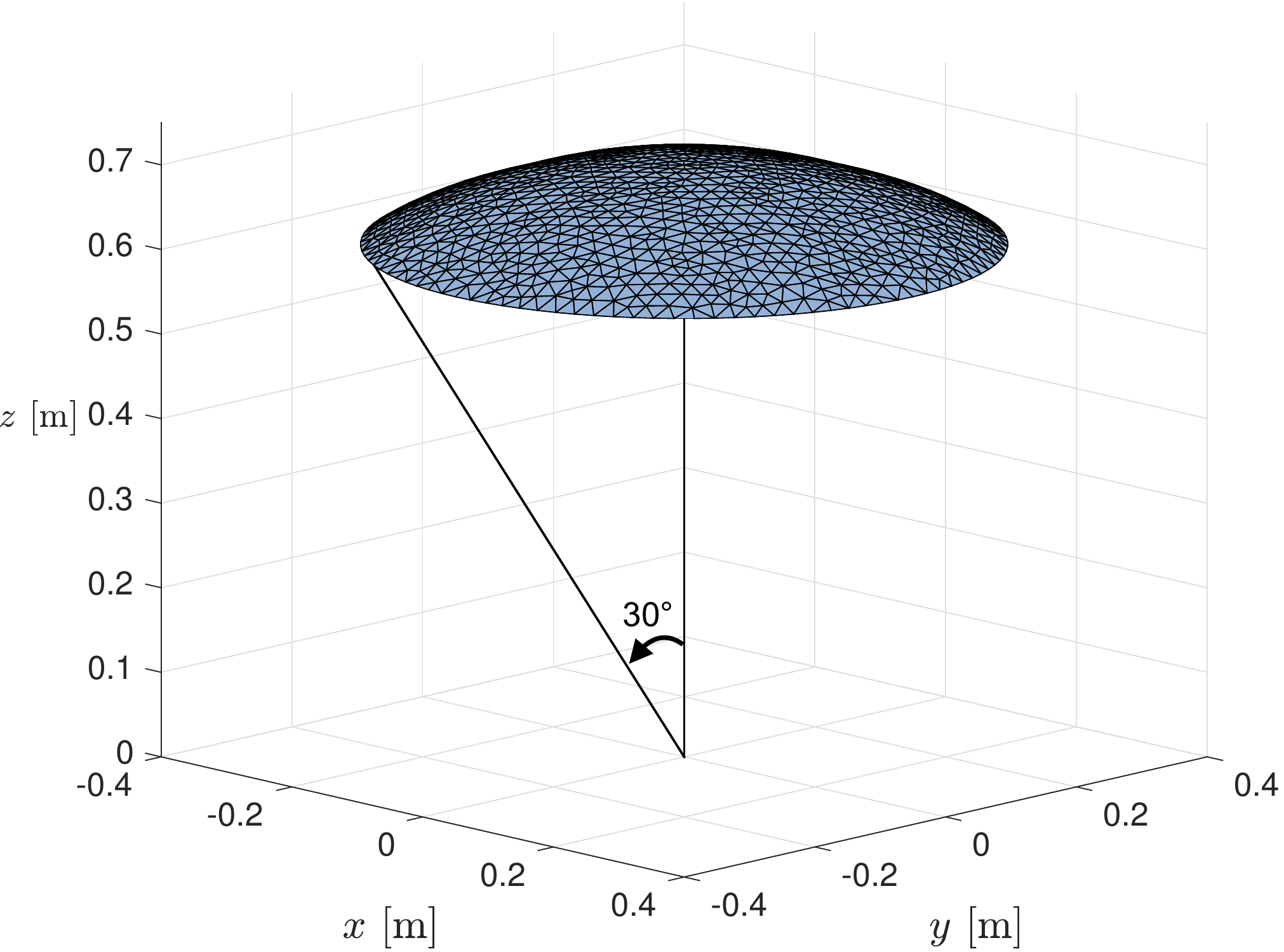}}
\caption{The spherical cap metasurface boundary $\Sigma$ used for the 3D example, which extends from $\theta=0^{\circ}$ to $\theta=30^{\circ}$ at a radius of 0.70 m. The equivalent currents are reconstructed on this surface.}
\label{fig:sphericalCap}
\end{figure}

\begin{figure}[t!]
     \centerline{\includegraphics[width=3.4in]{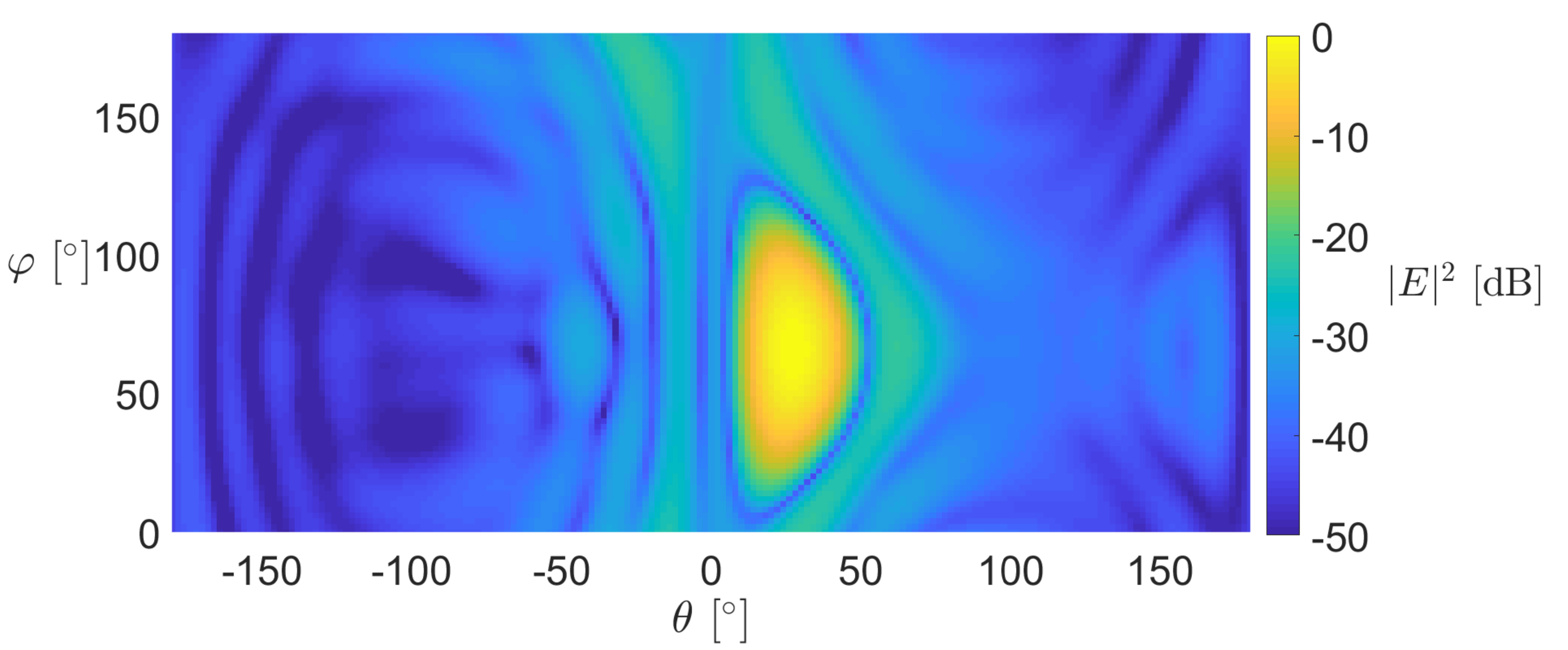}}
	\caption{The normalized far-field power pattern produced by the equivalent currents obtained by the inverse source algorithm for the desired specifications in Table~\ref{tab:3Dspecs}.}
         \label{fig:3DFF}
\end{figure}

\begin{figure}[t!]
         \centering
         \subfloat[]{\includegraphics[width=3.2in]{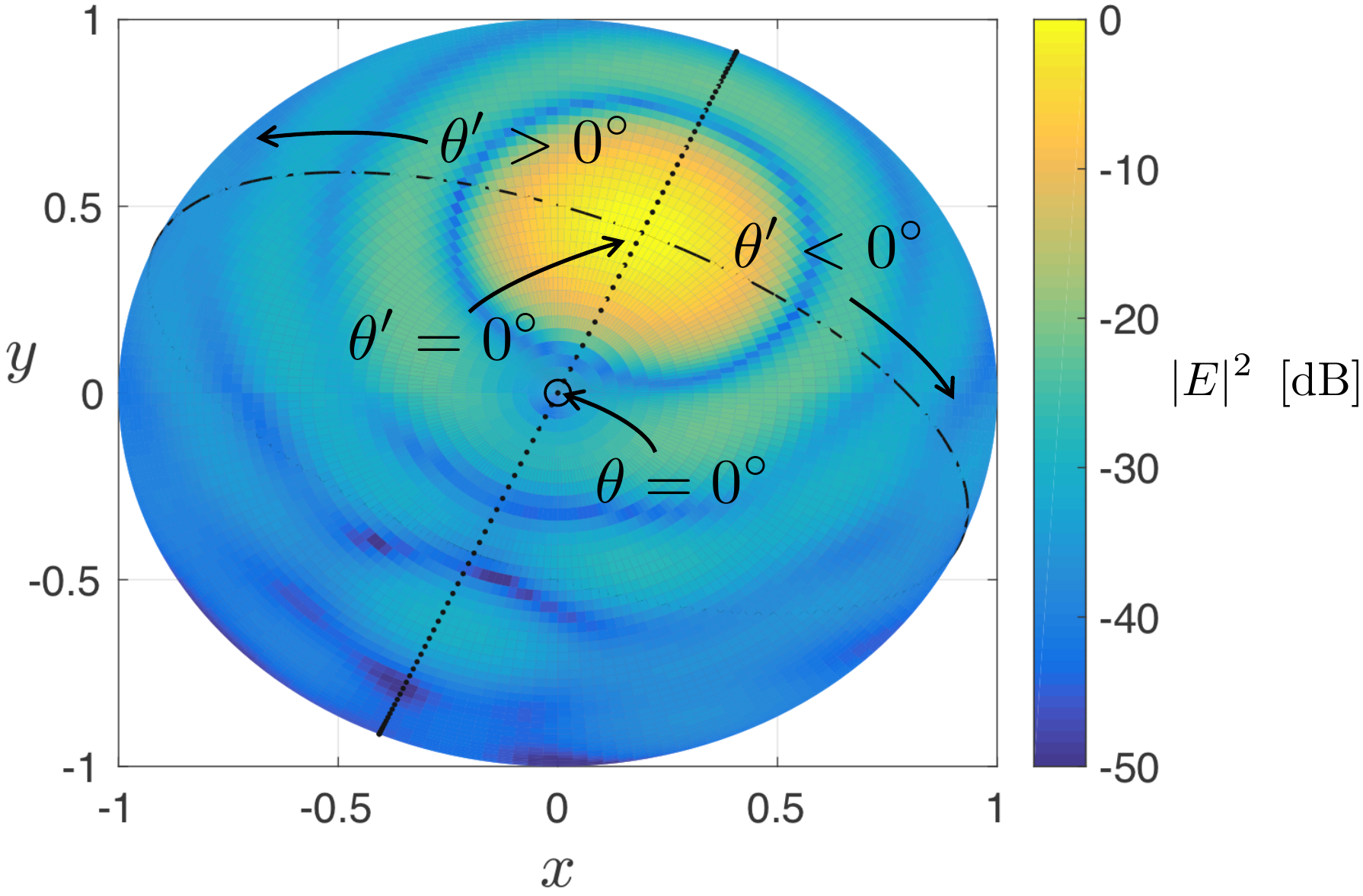}}\\
         \subfloat[]{\includegraphics[width=3.4in]{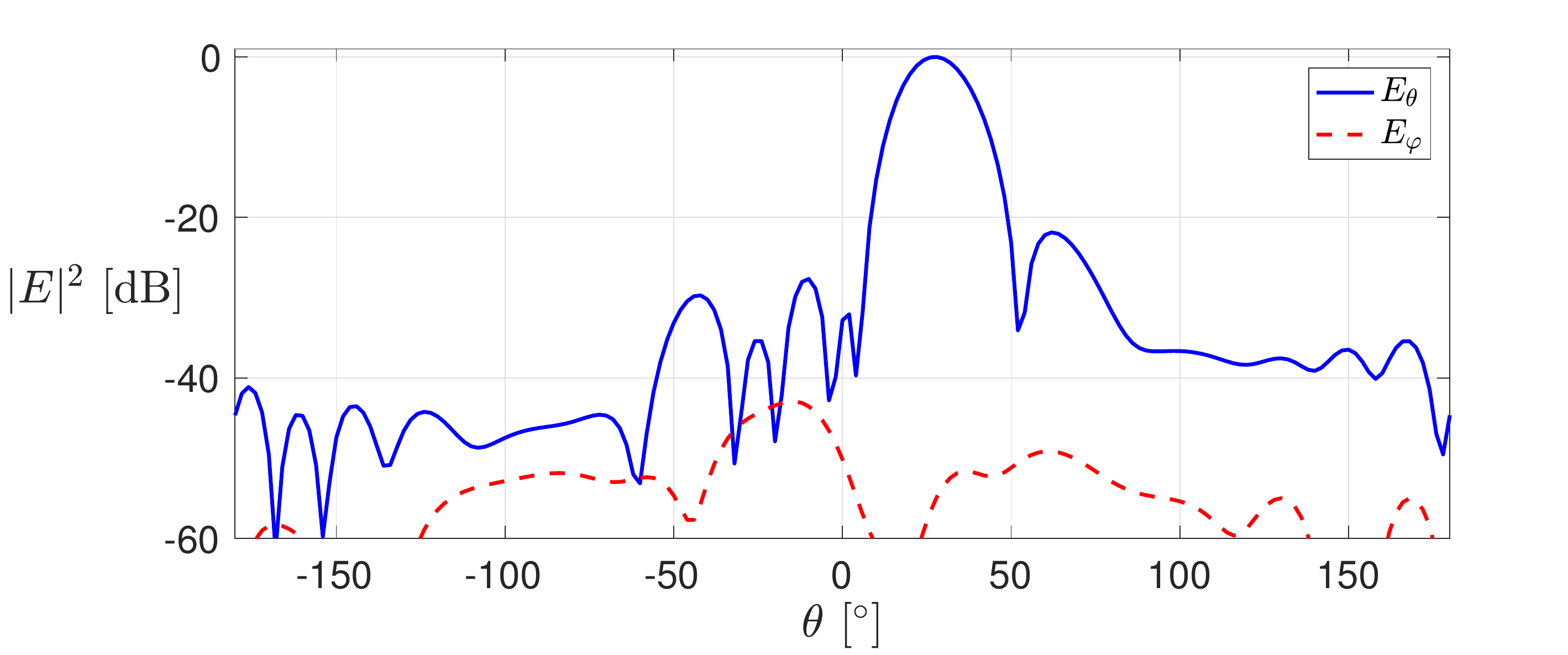}}\\
         \subfloat[]{\includegraphics[width=3.4in]{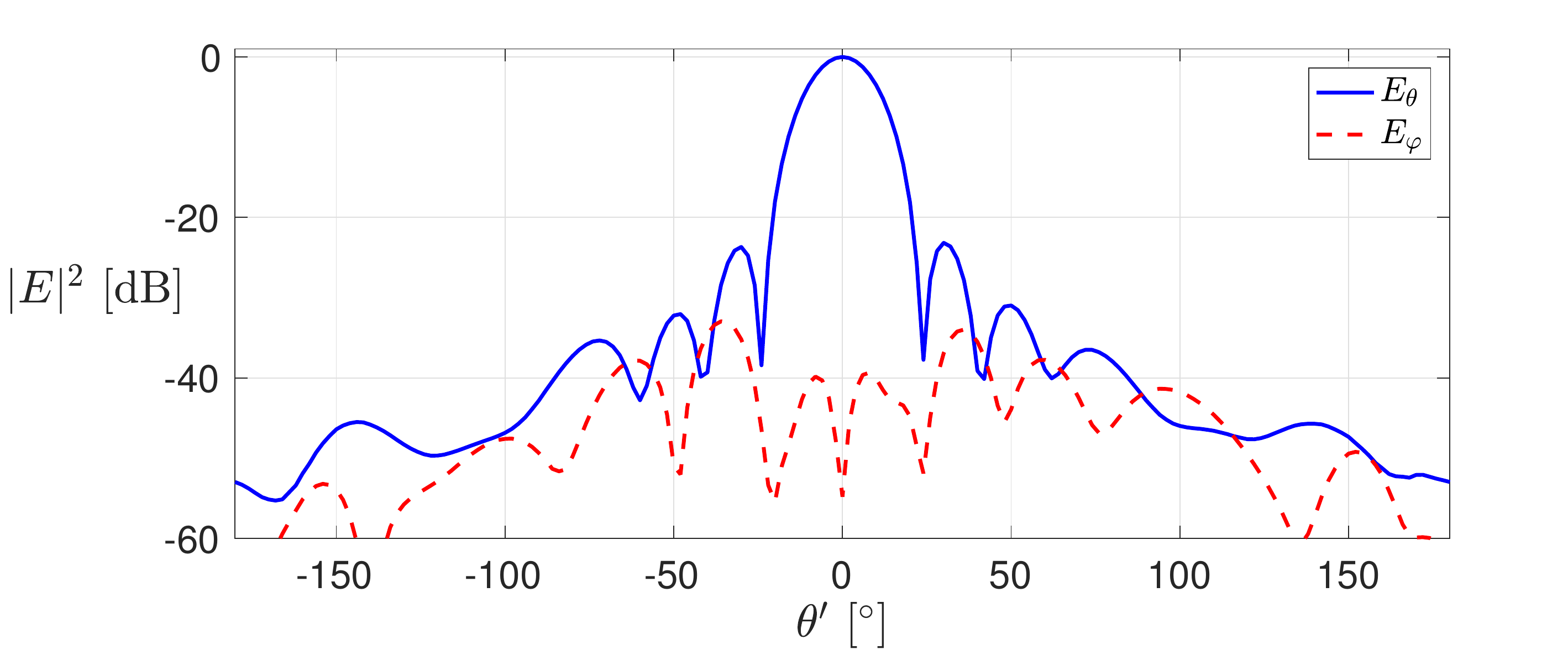}}
	\caption{The E-plane (b) and H-plane (c) cuts of the far-field pattern obtained by the inverse source algorithm for the desired specifications in Table~\ref{tab:3Dspecs}. For clarification, the FF power pattern is shown in (a) over a normalized sphere as viewed from the positive $z$ axis ($\theta = 0^{\circ}$), with the E-plane cut represented by the dotted line and the H-plane cut represented by the dash-dotted line. To accommodate plotting of the H-plane cut, a new angle denoted by $\theta '$ has been introduced such that $\theta ' = 0^{\circ}$ coincides with $(\theta = 28^{\circ}, \varphi = 66^{\circ})$ as shown in (a) and used as the horizontal axis in (c).}
         \label{fig:2DCuts}
\end{figure}

To display the generality of the proposed method, we now present an example with desired far-field performance criteria in 3D. The metasurface in this case is a spherical cap that extends from $\theta = 0$ to $\theta = 30^{\circ}$ at a radius of 0.70 m, as shown in Figure~\ref{fig:sphericalCap}. Love's condition is enforced on an inward-offset surface of the same shape but at a radius of 0.68 m, which results in a separation of approximately $\lambda/10$ between the two surfaces at the operating frequency of 1.7~GHz. The reconstruction surface $\Sigma$ in this example (i.e., the spherical cap) is discretized into triangular mesh elements and the equivalent currents are represented using Rao-Wilton-Glisson (RWG) basis functions~\cite{Rao1982Electromagnetic}. 
The desired FF performance criteria are shown in Table~\ref{tab:3Dspecs}. These specifications are enforced over a (partial) spherical domain of radius $500\lambda$ and an angular resolution of $1^{\circ}$. 
\begin{table}[t!]  
\renewcommand{\arraystretch}{1.3}
\caption{Desired Far-field (FF) Specifications -- 3D Example}
\label{tab:3Dspecs}
  \centering
  \begin{tabular}{l | l}
   \textbf{Specifications} & \textbf{Main Beam} \\ \hline
Direction & $\theta = 28^\circ, \varphi = 66^{\circ}$ \\ 
HPBW &$20^{\circ}$  \\
Nulls (relative to the main beam) &  $24^{\circ}$ at $-60$~dB \\
Polarization & Main beam linearly polarized \\&~~ in $\hat{\theta}$ direction
\end{tabular}
\end{table}
The inverse source algorithm is applied to~(\ref{eq:specfunctional}) to find the equivalent currents from the desired specifications. Once these currents are reconstructed, their power pattern is simulated using a 3D EFIE solver, which is shown in Figures~\ref{fig:3DFF} and~\ref{fig:2DCuts}. As can be seen, the main beam produced by the equivalent currents is along $(\theta=28^{\circ},\varphi=66^{\circ})$ as desired, and nearly satisfies the beamwidth requirement with a HPBW of $17.8^{\circ}$. Also, the first null is $24^\circ$ away from the main beam in a symmetric fashion. Moreover, within the main beam, the $\hat{\theta}$ component of the electric field is a \textit{minimum} of $31.5$~dB higher than the undesired $\hat{\varphi}$ polarization. In summary, all of the specifications have been met with only a minor deviation in the HPBW.

\section{Limitations}
\label{sec:limitations}
While the presented method has the potential to allow for more practical, non-idealistic metasurface designs, it is important to note the limitations of the current implementation. The main limitation arises from the fact that local power conservation is not enforced at any point during the design procedure, necessarily resulting in lossy/active elements. Future work will be focused on enforcing local power conservation during the inversion process, made possible by prior knowledge of the incident field. Furthermore, supporting such a transformation without loss or gain requires extra degrees of freedom, either by allowing for nonreciprocity or bianisotropy~\cite{chen2018theory,lavigne2018susceptibility} (e.g., by introducing magnetoelectric coupling). The reduction of the degrees of freedom from enforcing local power conservation will limit the solution space, and will most likely result in a worse fitting of the design specifications. 

Another limitation arises from the nature of the inverse problem, which is that an `appropriate' solution may not exist for a given set of design specifications and metasurface size and geometry. We currently do not have a method for determining if a `feasible' solution exists prior to the optimization process and therefore tuning of the metasurface parameters can be required in some cases.

\section{Conclusion}
A macroscopic metasurface design procedure has been presented that allows for a field transformation between incident, reflected, and transmitted fields. The desired transmitted field can be specified in a variety of ways, including complex field quantities, amplitude-only (phaseless) field data, or in terms of far-field performance criteria such as main beam directions, HPBWs, null locations, or polarization. Additionally, there are no restrictions on the locations at which the desired field is specified and the metasurface geometry, allowing for increased design flexibility. Several 2D examples were shown to demonstrate the capabilities of the proposed method for each of the different types of desired fields, and the designed metasurfaces were simulated using a FDFD-GSTC solver. The 3D example, although not verified through simulation of surface susceptibilities, was presented to demonstrate the most general form of the proposed method.

Many challenges remain, including ensuring the resulting metasurface susceptibilities are physically realizable. This may include incorporating restrictions into the inversion to ensure that the resulting elements are passive and lossless (as mentioned in Section~\ref{sec:limitations}), while also reducing high spatial variations. These restrictions can, for example, be implemented by adding extra terms to the data misfit cost functional. Similarly, we have not investigated how we can take advantage of the non-uniqueness of the inverse source problem to yield solutions that are more desirable. Lastly, it is possible to specify a desired field that cannot be produced by the metasurface. With this in mind, a procedure that can determine whether or not the desired field can be supported by the metasurface geometry would be a useful future addition.

%

\section*{Acknowledgment}

The authors would like to thank the Canadian Microelectronics Corp. for the provision of ANSYS HFSS and Mr. Nozhan Bayat for his assistance with the simulation studies. The financial support of the Natural Sciences and Engineering Research Council (NSERC) of Canada and the Canada Research Chair (CRC) Program is also acknowledged.

%
%

\bibliographystyle{IEEEtran}
\bibliography{IEEEabrv,MSDesign_TAP2019}
\vskip -1pt plus -1fil
\begin{IEEEbiography}[{\includegraphics[width=1in,height=1.25in,clip,keepaspectratio]{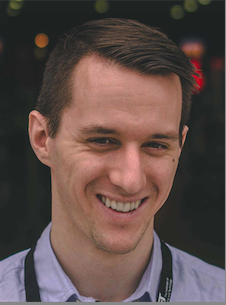}}]{Trevor Brown}
(S'12) received the B.Sc. and M.Sc. degrees in electrical engineering from the University of Manitoba, Winnipeg, MB, Canada, in 2014 and 2016, respectively, where he is currently pursuing the Ph.D. degree in electrical engineering.

His current research interests include applied and computational electromagnetics, electromagnetic metasurface design, inverse problems, optimization techniques, and phaseless near-field antenna measurement techniques.

Mr. Brown received the IEEE Antennas and Propagation Society Eugene F. Knott Memorial Pre-Doctoral Research Award in 2015. He held the Canada Graduate Scholarship (master's) from the Canadian National Sciences and Engineering Research Council (NSERC) from 2014 to 2016. He has also held the NSERC Postgraduate Scholarship since 2016. He has been serving as the Secretary of the IEEE Winnipeg Section since 2015. 
\end{IEEEbiography}

\vskip -1pt plus -1fil
\begin{IEEEbiography}[{\includegraphics[width=1in,height=1.25in,clip,keepaspectratio]{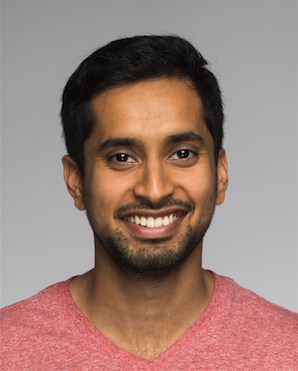}}]{Chaitanya Narendra}
(S'14) received a B.Sc. and M.Sc. in electrical engineering from the University of Manitoba, Winnipeg, MB, Canada, in 2014 and 2016, respectively. He is currently pursuing a Ph.D. in the same institution.

His research interests include electromagnetic inverse source and scattering problems, antenna design, and metasurface design.

Mr. Narendra was the recipient of the IEEE Eugene F. Knott Memorial Pre-Doctoral Research Award in 2015. He also held the Canadian National Sciences and Engineering Research Council (NSERC) Canada Graduate Scholarship (master's) from 2015-2016 and has held the NSERC Alexander Graham Bell Canada Graduate Scholarship since 2017. He started serving as the Treasurer of the IEEE Winnipeg Section in 2019.
\end{IEEEbiography}

\vskip 0pt plus -1fil
\begin{IEEEbiography}[{\includegraphics[width=1in,height=1.25in,clip,keepaspectratio]{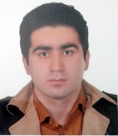}}]{Yousef Vahabzadeh}
(S'18) was born in Ardabil, Iran, on May 10, 1989. He received the master's degree in communication engineering, microwave and optics, from Sharif University of Technology, Iran, in February 2014. His M.Sc. project was on the design and characterization of a rectangular waveguide slot array antenna with tunable pattern using capacitive elements and was performed under supervision of Dr. Behzad Rejaei. He received the Ph.D. degree in July 2019 from \'{E}cole Polytechnique de Montr\'{e}al under the supervision of Prof. C. Caloz, where he worked on the design, computational analysis, and experimental development of metasurfaces.

He has been a postdoctoral researcher with Prof. Ke Wu at \'{E}cole Polytechnique de Montr\'{e}al since August 2019, where he is currently developing a harmonic radar system for vital sign detection. His current research interests include metamaterials and metasurfaces, computational electromagnetics, passive and active microwave and RF circuits, and antenna design.
\end{IEEEbiography}

\begin{IEEEbiography}[{\includegraphics[width=1in,height=1.25in,clip,keepaspectratio]{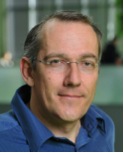}}]{Christophe Caloz} (F'10)
received the Dipl\^{o}me d'Ing\'{e}nieur en \'{E}lectricit\'{e} and the Ph.D. degree from \'{E}cole Polytechnique F\'{e}d\'{e}rale de Lausanne (EPFL), Switzerland, in 1995 and 2000, respectively.

From 2001 to 2004, he was a Postdoctoral Research Fellow at the Microwave Electronics Laboratory, University of California at Los Angeles (UCLA). In June 2004, Dr. Caloz joined \'{E}cole Polytechnique of Montr\'{e}al, where he is currently a Full Professor, the holder of a Canada Research Chair (Tier 1) and the head of the Electromagnetics Research Group. He has recently accepted a special position as a Full Research Professor at KU Leuven, and he will join that university in January 2020. Dr. Caloz has authored and co-authored over 750 technical conference, letter, and journal papers, 17 books and book chapters, and he holds dozens of patents. His works have generated over 26,000 citations (h-index=65), and he has been a Thomson Reuters Highly Cited Researcher. His current research interests include all fields of theoretical, computational, and technological electromagnetics, with strong emphasis on emergent and multidisciplinary topics, such as metamaterials and metasurfaces, nanoelectromagnetics, space-time electrodynamics, thermal radiation management, exotic antenna systems, and real-time radio/photonic processing.

Dr. Caloz was a Member of the Microwave Theory and Techniques Society (MTT-S) Technical Committees MTT-15 (Microwave Field Theory) and MTT-25 (RF Nanotechnology), a Speaker of the MTT-15 Speaker Bureau, the Chair of the Commission D (Electronics and Photonics) of the Canadian Union de Radio Science Internationale (URSI), and an MTT-S representative at the IEEE Nanotechnology Council (NTC). In 2009, he co-founded the company ScisWave (now Tembo Networks). He received several awards, including the UCLA Chancellor's Award for Post-doctoral Research in 2004, the MTT-S Outstanding Young Engineer Award in 2007, the E.W.R. Steacie Memorial Fellowship in 2013, the Prix Urgel-Archambault in 2013, the Killam Fellowship in 2016, and many best paper awards with his students at international conferences. He has been an IEEE Fellow since 2010, an IEEE Distinguished Lecturer for the Antennas and Propagation Society (AP-S) since 2014, and a Fellow of the Canadian Academy of Engineering since 2016. He was an Associate Editor of the Transactions on Antennas and Propagation of AP-S in from 2015 to 2017. In 2014, Dr. Caloz was elected as a member of the Administrative Committee of AP-S. He was also a Distinguished Adjunct Professor at King Abdulaziz University (KAU), Saudi Arabia, from May 2014 to November 2015.
\end{IEEEbiography}

\begin{IEEEbiography}[{\includegraphics[width=1in,height=1.25in,clip,keepaspectratio]{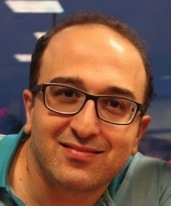}}]{Puyan Mojabi}
(S'09--M'10) Puyan Mojabi received the B.Sc. degree from the University of Tehran, Tehran, Iran, in 2002, the M.Sc. degree from Iran University of Science and Technology, Tehran, Iran, in 2004, and the Ph.D. degree from the University of Manitoba, Winnipeg, MB, Canada, in 2010, all in Electrical Engineering.

He is currently an Associate Professor and a Canada Research Chair (Tier 2) in the Department of Electrical and Computer Engineering at the University of Manitoba, and is a registered Professional Engineer in Manitoba, Canada. His current research is focused on electromagnetic inversion for characterization and design with applications in the areas of imaging, remote sensing, and antenna measurements and design.

Dr. Mojabi is a recipient of the University of Manitoba's Falconer Emerging Researcher Rh Award for Outstanding Contributions to Scholarship and Research in the Applied Sciences category and two Excellence in Teaching Awards from the University of Manitoba's Faculty of Engineering as well as a University of Manitoba Graduate Students Association Teaching Award. He has also received three Young Scientist Awards from the International Union of Radio Science (URSI), and is currently serving as an Early Career Representative of URSI's Commission K, and as the Chair of the IEEE Winnipeg Waves Chapter. 
\end{IEEEbiography}

\end{document}